\documentclass[preprint2]{aastex} 
\usepackage{savesym}
\savesymbol{tablenum}
\usepackage[output-decimal-marker={.}, exponent-product=\cdot, per-mode=fraction]{siunitx}
\restoresymbol{SIX}{tablenum}

\usepackage{amsmath, amssymb, bm}
\usepackage{tabularx}
\usepackage{multirow}
\usepackage{lscape}

\newcommand{\abs}[1]{\left| #1 \right|}

\shorttitle{Characteristics of Low-Latitude Coronal Holes}
\shortauthors{Hofmeister et al.}

\begin{document}

\title{Characteristics of Low-Latitude Coronal Holes near the Maximum of Solar Cycle 24}

\author{Stefan J. Hofmeister\altaffilmark{1}, Astrid Veronig\altaffilmark{1}, Martin A. Reiss\altaffilmark{1}, Manuela Temmer\altaffilmark{1}, Susanne Vennerstrom\altaffilmark{2}, Bojan Vr\v{s}nak\altaffilmark{3}, Bernd Heber\altaffilmark{4}}

\altaffiltext{1}{University of Graz, Institute of Physics, IGAM-Kanzelh\"ohe Observatory, Graz, Austria; \url{stefan.hofmeister@uni-graz.at}}
\altaffiltext{2}{National Space Institute, DTU Space, Denmark}
\altaffiltext{3}{Hvar Observatory, Faculty of Geodesy, Zagreb, Croatia}
\altaffiltext{4}{Universit\"at Kiel, Institut f\"ur Experimentelle und Angewandte Physik, Kiel, Germany}

\begin{abstract}
We investigate the statistics of 288 low-latitude coronal holes extracted from SDO/AIA-193 filtergrams over the time range 2011/01/01 to 2013/12/31. 
We analyse the distribution of characteristic coronal hole properties, such as the areas, mean AIA-193 intensities, and mean magnetic field densities, the local distribution of the SDO/AIA-193 intensity and the magnetic field within the
coronal holes, and the distribution of magnetic flux tubes in coronal holes. 
We find that the mean magnetic field density of all coronal holes under study is \SI{3.0 \pm 1.6}{G}, and the percentage of unbalanced magnetic flux is \SI{49 \pm 16}{\percent}. The mean magnetic field density, the mean unsigned magnetic field density, and the percentage of unbalanced magnetic flux of coronal holes depend strongly pairwise on each other, with correlation coefficients $cc > 0.92$.
Furthermore, we find that the unbalanced magnetic flux of the coronal holes is predominantly concentrated in magnetic flux tubes: \SI{38}{\percent} (\SI{81}{\percent}) of the unbalanced magnetic flux of coronal holes arises from only \SI{1}{\percent} (\SI{10}{\percent}) of the coronal hole area, clustered in magnetic flux tubes with field strengths $> \SI{50}{G}\ (\SI{10}{G})$.
The average magnetic field density and the unbalanced magnetic flux derived from the magnetic flux tubes correlate with the mean magnetic field density and the unbalanced magnetic flux of the overall coronal hole ($cc > 0.93$). These findings give evidence that the overall magnetic characteristics of coronal holes are governed by the characteristics of the magnetic flux tubes.\\[1cm]
\end{abstract}

\section{INTRODUCTION}

Coronal holes are the lowest density regions in the solar atmosphere. Their low density and temperature as compared to the ambient corona results in a reduced emission in the extreme-ultraviolet (EUV) and soft X-ray (SXR). Therefore, they appear as dark structures in EUV images, and are often identified by intensity based thresholding techniques \citep[e.g.\ ][]{krista2009, rotter2012}. 

In addition, coronal holes are typically characterized by a large amount of ``open'' magnetic flux, i.e., magnetic flux which closes at far distances to the Sun. The open magnetic flux of coronal holes expands rapidly above the bottom of coronal holes due to the low ambient pressure and the low plasma-to-magnetic pressure ratio $\beta$. At a distance between about 1~R$_\odot$ and 10~R$_\odot$ above the solar-photosphere surface, plasma is accelerated along the open magnetic field lines to form high-speed solar wind streams, i.e., supersonic plasma flows which transverse our solar system \citep[e.g.\  review by ][]{kohl2006}. High-speed solar wind streams are the major cause of minor and medium geomagnetic storms at Earth.

The peak velocity of high-speed streams measured at L1 is correlated to the inverse expansion factor of the rapidly expanding magnetic flux tubes of coronal holes evaluated between the lower boundary of coronal holes and the source-surface at 2.5~R$_\odot$ as derived by potential-field source-surface (PFSS) extrapolations \citep{wang1990}. In addition, the peak velocity of high-speed streams correlates with the ratio of the mean magnetic field density of the photosphere below a CH to its flux tube expansion factor, and with the area of the coronal hole \citep{nolte1976, kojima2007}.

For these reasons, the magnetic field distribution in coronal holes is of high interest.
\citet{harvey1982} and \citet{wang2009} studied the evolution of the mean magnetic field densities of coronal holes over the solar cycle. They found that the mean magnetic field densities at solar minimum are about \SI{5}{G}, respectively \SIrange{0.3}{7.2}{G}, and that they are 3-5 times higher at solar maximum.
\citet{bohlin1978} studied the magnetic characteristics over the life time of coronal holes. They report that the mean magnetic field densities of coronal holes decrease with their age.

\cite{wiegelmann2004} studied 12 well pronounced coronal holes in comparison to quiet Sun regions. They report that the relative amount of the total signed magnetic flux to the total unsigned magnetic flux in their coronal holes under study is \SI{77 \pm 14}{\percent}, i.e.\ that most of the magnetic flux in coronal holes is open. However, these authors also report that a significant number of closed magnetic loops exists within coronal holes. They have shown that short and low-lying coronal loops are almost as abundant in coronal holes as in quiet Sun regions, whereas high and long loops are very rare. 
PFSS extrapolations confirmed that the global magnetic field in coronal holes is predominantly unipolar, open, and expands super-radially with height, whereby the flux tube expansion factor is highest at the boundary of coronal holes and smallest near the center \citep[e.g.\ ][]{altschuler1972, levine1977a, levine1977b, wiegelmann2004}.
 \cite{fainshtein1994} reported that the flux tube expansion factor derived at the source-surface, the mean magnetic field density at the base, and the area at the base of the coronal hole depend on each other. Furthermore, they found that the mean magnetic field density derived at the source-surface at 2.5~R$_\odot$ correlates well with the area at the base of coronal holes.
 
Correlating the force-free extrapolated magnetic field at different altitudes of coronal holes with Doppler-velocity maps derived from SI$^+$, C$^{3+}$ and Ne$^{7+}$ spectral lines, \citet{tu2005} found that the solar wind plasma flows out in magnetic funnels at heights between \SIrange{5}{20}{Mm} above the photosphere. 
 \cite{xia2004} showed that the outflow velocity is statistically higher at higher underlying photospheric magnetic field strengths.
Finally, \cite{hassler1999} showed by spectroscopic observations of the Ne$^{7+}$ line in a polar coronal hole that high-speed solar wind streams are rooted near the supergranulation boundaries in the chromospheric network lanes below coronal holes.

In this paper, we give a comprehensive statistical overview over 288 records of low-latitude coronal holes detected in EUV images in the time range 2011/01/01 to 2013/12/31, i.e.\ from the rising phase of solar cycle 24 up to the solar maximum. Note that solar maxima, as inferred from the sunspot numbers, tend to be double-peaked \citep{gnevyshev1963, gnevyshev1977}; the first peak of solar cycle 24 appeared in March, 2012 corresponding to the northern solar hemisphere and the second, slightly larger peak in April, 2014 corresponding to the southern hemisphere. Our dataset contains all the coronal holes detected at a cadence of one image per day with areas greater than \SI{e10}{km^2} and center of mass situated within \SI{30}{\degree} around the intersection of the solar equator and the central meridian. The coronal holes were allowed to expand beyond the \SI{30}{\degree} restriction, but equatorial extensions of polar coronal holes are not included.
 
We divided this paper into three main parts. The first part presents the distribution of characteristic observational properties of the low-latitude coronal holes, such as their areas, the mean EUV intensity, mean magnetic field densities, the mean unsigned magnetic field densities, the unbalanced magnetic flux, and the percentage unbalanced magnetic flux, i.e.\ the relative amount of the total signed to total unsigned magnetic flux. We investigate their correlations, and examine the difference in the statistics between small and large coronal holes. Most notably, we find a strong dependency between the mean magnetic field density, the mean unsigned magnetic field density, and the percentage unsigned magnetic flux of coronal holes.

In the second part, we investigate the distribution of the AIA-193 intensity and the magnetic field density within coronal holes, in particular the probability distribution of the AIA-193 intensity along a normalized cross-section of coronal holes, the magnetic polarity of coronal holes at various magnetic levels, and the amount of unbalanced magnetic flux that arises from various magnetic levels.
We find that the magnetic field distributions at various magnetic levels of a coronal hole always have the same polarity, and that most of the unbalanced magnetic flux arises from only a small percentage of its area.

In the third part, we present magnetic flux tubes as the origin of the unbalanced magnetic flux of coronal holes. We investigate the distribution of magnetic flux tubes within coronal holes, and relate the averaged properties of magnetic flux tubes within coronal holes to the overall magnetic properties of coronal holes. We find strong dependencies between the mean magnetic field density and unbalanced magnetic flux of the flux tubes, of the coronal hole region without the flux tubes, and of the magnetic properties of the overall coronal hole.

The paper is structured as follows: Section \ref{dataset} describes the dataset and the data reduction, Section \ref{analysis} the analysis methods. Section \ref{stchardis} 
studies the distribution of characteristic observational properties of the low-latitude coronal holes. Section \ref{pxdist} presents the local distribution of the EUV intensity and the magnetic field within the coronal holes. Section \ref{flxtb} studies the distribution and properties of magnetic flux tubes within the coronal holes. Section \ref{correlateftch} relates the averaged parameters of magnetic flux tubes to the overall magnetic properties of the coronal holes. Section \ref{discussion} discusses the results.

\section{DATASETS AND DATA REDUCTION}
\label{dataset}
We use EUV \SI{193}{\angstrom} filtergrams recorded by the Atmospheric Imaging Assembly (AIA) on-board of the Solar Dynamics Observatory (SDO), low noise line-of-sight magnetograms recorded by the Helioseismic and Magnetic Imager (HMI) on-board of SDO and provided by the Joint Science Operations Center (JSOC), and H$\alpha$ images from the Kanzelh\"ohe Observatory of Solar and Environmental Research.

To identify coronal holes in EUV images, we use AIA \SI{193}{\angstrom} filtergrams. Because of their high contrast between coronal holes and quiet Sun regions, these filtergrams are often used to identify coronal holes in EUV images by various segmentation techniques \citep[e.g.\ ][]{krista2009, rotter2012, verbeek2014, lowder2014, caplan2016, boucheron2016}.
The AIA \SI{193}{\angstrom} filtergrams observe emission from Fe XII ions in the coronal plasma at a temperature of about \SI{1.6}{MK} (peak response). 
Images are taken by a $4096 \times 4096$ pixel CCD camera, the spatial resolution is \SI{1.5}{arcsec} at a plate scale of \SI{0.6}{arcsec/pixel}  \citep{lemen2012}. 

To study the photospheric magnetic field below coronal holes, we use the HMI-los 720s magnetogram dataset. This magnetogram data product is a low noise line-of-sight magnetogram data product derived by the Vector Camera of the HMI instrument. We prefer this data product to the HMI vector magnetograms due to its much higher signal-to-noise ratio at weak magnetic fields. The Vector Camera derives the Stoke's parameters for the full solar disk from the photospheric Fe I line (\SI{6173}{\angstrom}). Ten sets of Stoke's parameters recorded at a cadence of \SI{135}{s} are then averaged to derive the line-of-sight magnetic field. The plate scale is \SI{0.505}{arcsec/pixel} at $4096 \times 4096$ pixels, the spatial resolution is \SI{0.91}{arcsec}, and the noise level is \SI{5}{G/pixel}. Note that the noise level primarily arises from photon noise, and therefore decreases with $\sqrt{n}$ by averaging over $n$ pixels \citep{scherrer2012, schou2012, couvidat2016}. 

To distinguish coronal holes from filaments, we use H$\alpha$ filtergrams taken by Kanzelh\"ohe Observatory. The H$\alpha$ filtergrams observe emission from neutral hydrogen in the chromosphere at the center of the H$\alpha$ line (\SI{6563}{\angstrom}) with a FWHM of \SI{0.7}{\angstrom}. The images are recorded by a $2048 \times 2048$ pixel CCD camera at a plate scale of \SI{1}{arcsec/pixel} \citep{potzi2015}.

All images were normalized to an exposure time of \SI{1}{s}, rotated to north up, and interpolated to a spatial sampling of \SI{2.4}{arcsec/pixel} at $1024 \times 1024$ pixels. The interpolation to $1024 \times 1024$ pixels was performed under the condition of conservation of flux, and reduces the noise level of the line-of-sight magnetograms to about \SI{1.25}{G} per pixel. Due to the different recording times of the AIA-193, HMI-los, and H$\alpha$ images, the magnetograms and H$\alpha$ images were co-aligned assuming rigid rotation of the Sun to match the recording time of the AIA-193 images. The maximum time of de-rotation was \SI{6}{min} for magnetograms and \SI{12}{h} for H$\alpha$ images. Since H$\alpha$ images are only used for the visual identification of filaments, the error in the de-rotation of H$\alpha$ images by assuming rigid rotation over \SI{12}{h} can be neglected.

\section{ANALYSIS}
\label{analysis}

\subsection{Extraction of the Coronal Holes}

Two species of structures appear dark in EUV images: coronal holes and filaments.
Coronal holes are the lowest density structures in the Sun's corona at comparably low coronal temperatures of about \SI{1}{MK}. Therefore they appear as dark structures in EUV images, but are not visible in H$\alpha$ images. In contrast, filaments have a high density, a chromospheric temperature of about \SI{e4}{K}, extending from the Sun's chromosphere to the lower corona. Filaments appear dark in both EUV and H$\alpha$ images. 

To create our dataset, we first extracted all dark structures in EUV images for the 3-year-period under study from 2011/01/01 to 2013/12/31 with a cadence of one image per day. Then we visually classify and exclude filaments within the dark structures by comparison with H$\alpha$ images. The dataset is restricted by the availability of EUV images, H$\alpha$ images and magnetograms, and covers \SI{98}{\percent} of the time range under study.

We extracted the dark structures in the EUV images by an intensity-based thresholding algorithm based on \citet{rotter2012} and \citet{reiss2016}. 
The threshold was set to 0.35 times the median of the AIA-193 intensity of the solar disk.
The resulting full disk binary maps of dark structures were post-processed by morphological operators (erosion and dilation), whereby a median function with a kernel of 9 pixels was applied. This morphological operation adds small bright islands within coronal holes to the area of coronal holes, and removes very small dark structures, which are most probably neither coronal holes nor filaments, from the full-disk binary map \citep{rotter2012}.
In order to obtain the single structure binary maps the full disk binary maps were segmented. Two structures were recognized as individual structures as soon as they are separated by at least one pixel. Finally, the EUV images and magnetograms of the single structures were obtained by cutting the full-disk EUV images and magnetograms according to the single structure binary maps identified in the EUV. 

\begin{figure*}[tp]
\centering
\includegraphics[width = \textwidth]{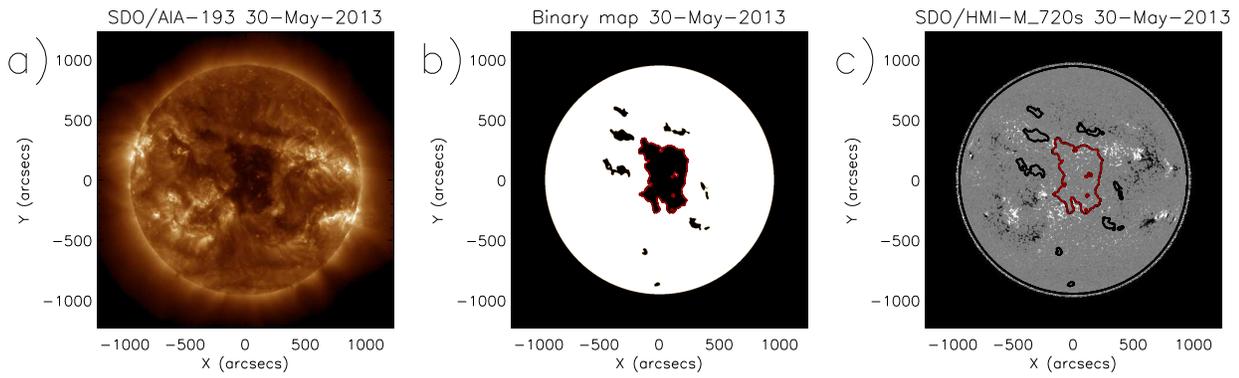}
\caption{a) SDO/AIA-193 sample filtergram of the Sun, taken on 05/29/2013. A large coronal hole and surrounding small coronal holes and filaments are visible as dark structures. b) Binary map of the identified structures. Only the red-outlined structure has an area of $> \SI{e10}{km^2}$ ($A = \SI{9.1e10}{km^2}$). c) SDO/ HMI-M\_{720s} line-of-sight magnetogram of the Sun. The contours of the binary map and the photospheric solar limb are overplotted. }
\label{statistics17}
\end{figure*}

For further analysis, we restricted the single structure dataset to structures with an area of more than $1\times 10^{10}$ km$^2$ corresponding to a minimum diameter of \SI{100}{Mm}, and with an angular distance of the structure's center of mass to the intersection of the solar equator and central meridian of less than \SI{30}{\degree} in order to reduce the effects of projection in the further analysis. Thus, we arrive at a resulting dataset of 343 structures. Figure \ref{statistics17} shows an example of the segmentation procedure and resulting coronal holes. 

The remaining single structures identified in EUV were then divided into coronal holes and filaments by visual comparison with the H$\alpha$ images. Since filament channels are sometimes visible in EUV before the corresponding channel is visible in H$\alpha$, i.e.\ before the channel is filled with dense cool plasma, we inspected the H$\alpha$ images within a time range of $\pm 2$ days.
If the full structure was visible in the H$\alpha$ images within a time range $\pm 2$ days, the structure was labelled as filament. If only a part of the structure was visible in H$\alpha$, the structure was excluded from further analysis. If the structure was not visible in H$\alpha$, but had an elongated shape and a balanced magnetic flux like we expect it from filaments, the structure was also excluded from further analysis. All structures remaining were labelled as coronal holes.

In total, our dataset includes 288 records of dark structures that were identified as coronal holes.
49 dark structures which were identified as filaments, and 6 dark structures which could not be uniquely assigned, were excluded from further analysis.

\subsection{Analysis of the Characteristic Properties of the Coronal Holes}
\label{analysischarpars}

We analysed the distribution of characteristic properties in the coronal hole dataset, i.e.\ the areas, latitudes of the center of mass, the mean EUV intensities, the mean value and skewness of the distribution of the radial magnetic field in the photosphere, and the magnetic flux.
 
We denote a pixel within a coronal hole as $i$, its projected area as $A_{i,proj}$, its AIA-193 intensity as $I_{193,i}$, its heliographic latitude as $\varphi_i$, its line-of-sight magnetic field density as $B_{i,los}$, and the total number of pixels within the coronal hole as $n$.

First, in order to avoid systematic errors due to the different latitudes of the coronal holes, we presume the magnetic field to be radial and correct each pixel for projection effects:
\begin{align}
&A_{i} = \frac{A_{i,proj}}{\cos \alpha_i}, \\
&B_{i} = \frac{B_{i,los}}{\cos \alpha_i},
\end{align}
where $\alpha_i$ is the heliographic angular distance of the pixel to the center of the solar disk as seen from SDO.

Then, for each coronal hole, we calculated the projection-corrected area $A$, the latitude of its center of mass $\varphi$, and the mean EUV-intensity in the \SI{193}{\angstrom} filtergrams $\bar{I}_{193}$.

Further, we calculated the projection-corrected unbalanced magnetic flux $\Phi$, the total unsigned magnetic flux $\Phi_\text{t}$, and the percentaged unbalanced magnetic flux $\Phi_\text{pu}$, by
\begin{align}
&\Phi =  \sum_i B_i A_i ,\\
&\Phi_\text{t} =  \sum_i \abs{ B_i A_i },  \\
&\Phi_\text{pu} = \frac{\Phi}{\Phi_\text{t}}.
\end{align}
The unbalanced magnetic flux gives the magnetic flux which is not closed \textit{within} the coronal hole. Under the standard assumption that only magnetic flux near the border of the coronal hole closes with neighbouring regions, the unbalanced magnetic flux is a reasonable estimate of the open magnetic flux.
The percentage unbalanced magnetic flux gives the percentage of magnetic flux which is not closed within the coronal hole. 

Finally, we calculated the projection-corrected mean magnetic field density $\bar{B}$ and the projection-corrected mean unsigned magnetic field density $\bar{B}_\text{us}$ in the photosphere below the coronal hole,
\begin{align}
&\bar{B} = \frac{\Phi}{A}, \\
&\bar{B}_\text{us} = \frac{\Phi_\text{t}}{A},
\end{align} 
and the skewness Skew($B$) of the magnetic field distribution. In our context the skewness is a measure of the distinctness of a dominant polarity.

\subsection{Analysis of the Pixel Distribution within the Coronal Holes}

Whereas in Section \ref{analysischarpars} we analyse the distribution of characteristic properties of coronal holes, here we analyse the distribution of AIA-193 and magnetogram pixels \textit{within} the coronal holes. To do so, we calculate for each coronal hole a normalized histogram. Then, we stack the normalized histograms of all coronal holes, whereby all histograms are weighted equally, i.e.\ small coronal holes get the same weight as large coronal holes. The projection of the stacked histograms gives us a median representation of the histograms and a $1\sigma$ and a $2\sigma$ range in which the histograms lie, i.e.\ a probability distribution of the histograms. This approach provides us with a characterisation of the average pixel distribution within all coronal holes under study. 

First we analysed the distribution of the EUV intensity within coronal holes. For each coronal hole we created a histogram of the pixels of the AIA-193 intensity $I_{193,i}$, and normalized the histograms by their total number of pixels to obtain for each coronal hole the percentage of pixels belonging to each intensity bin. Then we calculated separately from all coronal holes the median value and the $1\sigma$ and $2\sigma$ ranges of the percentages of pixels for each intensity bin.

Next we analysed the EUV intensities along the major axis of coronal holes. First we calculated the direction of the major axis by using a least-square fit on the longitude-latitude tupels of all pixels of the coronal hole. We calculated the length of all cross-section parallel to the derived direction to get the longest cross-section, which we denote as major axis. 
We define the width of the coronal hole as the number of pixels along the longest cut perpendicular to the major axis. The AIA-193 intensity along the major axis was then derived by averaging over \SI{\pm 1}{\percent} of the width of the coronal hole perpendicular to major axis. 
For each coronal hole we derived the AIA-193 intensity versus the position along the normalized major axis. We then calculated from all coronal holes the representative median AIA-193 intensity and the $1\sigma$ and $2\sigma$ ranges for the normalized positions along the major axis. 

Further we analysed the distribution of the magnetic field densities within coronal holes. Since the dominant polarity of coronal holes can be positive or negative, we transform the magnetic field densities $B_i$ within a coronal hole so that positive values always belong to the dominant polarity of the coronal hole:
\begin{equation}
B_+ = \begin{cases} \ \ B_i & \text{for } \bar{B} \ge 0 \\ -B_i & \text{for } \bar{B} < 0 \end{cases}.
\end{equation}
For each coronal hole we created a histogram of $B_+$.

Next, we analysed the distinctiveness of the dominant polarity at various magnetic field density levels within coronal holes. We define the magnetic flux arising from a given signed magnetic field density level $l$ as 
\begin{equation}
\Phi_\text{lv}(l) = \sum_{\abs{-B_+ + l} < \SI{1}{G} } B_+ A_i,
\end{equation}
where $\abs{-B_+ +l} < \SI{1}{G}$ corresponds to the binning of the magnetic field density values $B_+$. Then, we denote the fraction of the magnetic flux arising from a given magnetic field density level $l$ which has the same polarity as the overall coronal hole as the polarity dominance $\Phi_\text{pd}(l)$ of the magnetic field density level $l$,
\begin{equation}
\Phi_\text{pd}(l) = \frac{\abs{\Phi_\text{lv}(l)}}{\abs{\Phi_\text{lv}(l)} + \abs{\Phi_\text{lv}(-l)}}.
\end{equation}
A polarity dominance of $\Phi_\text{pd}(l)=0.5$ means that the magnetic flux arising from the magnetic field density level $l$ is balanced, whereas a polarity dominance of $\Phi_\text{pd}(l)=1$ ($\Phi_\text{pd}(l)=0$) corresponds to an unipolar magnetic flux arising from the magnetic field density level $l$ with the same (opposite) polarity as the overall coronal hole.

Finally, we analysed the fraction of the unbalanced magnetic flux $\Phi$ of a coronal hole that arises from all pixels above a given threshold magnetic field density $b$, $\Phi_\text{fu}(b)$:
\begin{equation}
\Phi_\text{fu}(b) = \frac{\sum_{\abs{B_+} \ge b} B_+ A_i }{\Phi}.
\end{equation}

\subsection{Analysis of the Distribution of Magnetic Flux Tubes within the Coronal Holes}
\label{analysisfluxtubes}

We investigated the characteristics of magnetic flux tubes within each coronal hole.
We extracted the magnetic flux tubes by a thresholding technique on the magnetograms of the coronal holes at different global thresholds of $thr = i \cdot$\SI{10}{G}, $i \in \{1, 2, ..., 8\}$: First all pixels in the magnetogram with $\abs{B_i} < thr$ were set to Not a Number (NaN). Each island remaining within the magnetograms is defined as a magnetic flux tube. The resulting magnetogram was segmented, whereby two flux tubes were identified as individual flux tubes as soon as they are separated by at least one pixel. Note that the flux tubes extracted and their parameters derived depend on the magnetic extraction threshold $thr$. 

For each flux tube extracted at a threshold $thr$ we calculated the following parameters: its projection-corrected area $A_{\circ,thr}$, its projection-corrected mean magnetic field density $\bar{B}_{\circ,thr}$, and its projection-corrected magnetic flux $\Phi_ {\circ,thr}$ (cf. Sect. \ref{analysischarpars}).

For each coronal hole we counted the number of flux tubes $\tilde{N}_{\circ,thr}$. We calculated the total area $\tilde{A}_{\circ,thr}$ of the coronal hole that is covered by flux tubes, their summed unbalanced magnetic flux $\tilde{\Phi}_{\circ,thr}$, their summed total unsigned magnetic flux $\tilde{\Phi}_{\circ,t,thr}$, and their averaged magnetic field density $\tilde{B}_{\circ,thr}$, by summing (averaging) over all pixels which belong to magnetic flux tubes extracted at a given threshold $thr$:
\begin{align}
&\tilde{A}_{\circ,thr} =  \sum_{\abs{B_i} \ge thr} A_i \\
&\tilde{\Phi}_{\circ,thr} =  \sum_{\abs{B_i} \ge thr} B_i A_i \\
&\tilde{\Phi}_{\circ,t,thr} = \sum_{\abs{B_i} \ge thr} \abs{B_i A_i}, \\
&\tilde{B}_{\circ,thr} = \frac{1}{\tilde{A}_{\circ,thr}} \sum_{\abs{B_i} \ge thr} B_i A_i.
\end{align}

In addition, we calculated the mean magnetic field density of the coronal hole without the flux tubes, which we denote as ``quiet'' coronal hole regions:
\begin{equation}
\bar{B}_{\text{qu,}thr} =  \frac{1}{\left( A - \tilde{A}_{\circ,thr} \right)} \sum_{\abs{B_i} < thr} B_i A_i,
\end{equation}
where A is the area of the coronal hole.

Finally, for each coronal hole we estimated the fraction of the magnetic flux arising from all flux tubes which is unbalanced, i.e.\ the percentage unbalanced magnetic flux of the flux tubes 
\begin{equation}
\tilde{\Phi}_{\text{pu,}thr} = \frac{\tilde{\Phi}_{\circ,thr}}{\tilde{\Phi}_{\circ,t,thr}}.
\end{equation}
Note that under condition of similar absolute unbalanced magnetic fluxes $\abs{\Phi_{\circ,thr}}$ of the flux tubes, $\tilde{\Phi}_{\text{pu,}thr}$ is also an estimate on the number of flux tubes which are open and on the number of flux tubes which are closed loops. 

An overview on all properties defined is given in Table \ref{DefinitionAttributes}.

\section{CHARACTERISTIC PROPERTIES OF LOW-LATITUDE CORONAL HOLES}
\label{stchardis}

In this section, we present the statistics of characteristic properties of the 288 records of low-latitude coronal holes from 2011/01/01 to 2013/12/31. 
In particular, we present the area, the latitude, the mean AIA-193 intensity, the mean magnetic field density, the mean unsigned magnetic field density, the skewness of magnetic field distribution, the unbalanced magnetic flux, and the percentage unbalanced magnetic flux. In addition, we investigate the correlations between characteristic properties, and the differences in the distributions between small and large coronal holes.

\subsection{Areas, Latitudes, and Polarities}
\begin{figure*}[tp]
\centering
\includegraphics[width = .5\textwidth]{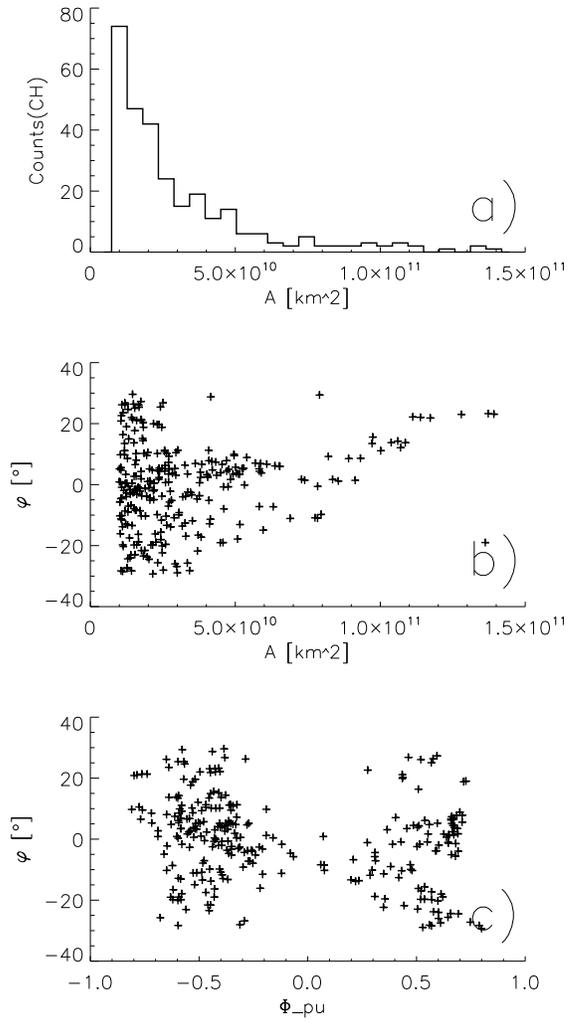}
\caption{Distribution of the areas of the coronal holes (a). Scatter plots of the heliographic latitude of the center of mass of coronal holes versus their areas (b) and versus their percentaged unbalanced magnetic flux (c).}
\label{statistics1}
\end{figure*}

Figure \ref{statistics1} gives an overview over the dataset of the low-latitude coronal holes, i.e.\ the distribution of the areas, latitudes and polarities of all coronal holes under study.

Figure \ref{statistics1}a shows the distribution of the deprojected areas $A$ of the coronal holes. The median area is \SI{2.39e10}{km^2}. Only \SI{18}{\percent} of all coronal holes under study had areas $A > \SI{5e10}{km^2}$. The largest coronal hole detected had an area of \SI{1.4e11}{km^2}, i.e.\ covering \SI{2.25}{\percent} of the visible solar hemisphere.

Figure \ref{statistics1}b shows the heliographic latitudes $\varphi$ of the center of mass of coronal holes versus their areas $A$. In the time range under study large coronal holes appeared predominantly on the northern hemisphere, whereas small and medium sized coronal holes appeared at both hemispheres. 

Figure \ref{statistics1}c shows the latitudes $\varphi$ of the center of mass of coronal holes versus their percentaged unbalanced magnetic flux $\Phi_\text{pu}$, which corresponds to their magnetic polarity. In the time range under study, i.e.\ the rising phase of solar cycle 24 up to solar maximum, coronal holes appeared with both polarities in each of the two hemispheres at almost all times. This can be explained by the reversal of the magnetic polarity of polar coronal holes during the time range under study. Coronal holes with a small percentage of unbalanced magnetic flux, i.e.\ a weakly pronounced polarity, only appeared near the solar equator, with a medium percentage of unbalanced magnetic flux at all latitudes, and with a high percentage of unbalanced magnetic flux only at some distance to the solar equator. 
Note that this plot is not symmetric to the solar equator, but to \SI{\approx 5}{\degree} south.

\subsection{Mean AIA-193 Intensity}
\begin{figure*}[tp]
\centering
\includegraphics[width = .5\textwidth]{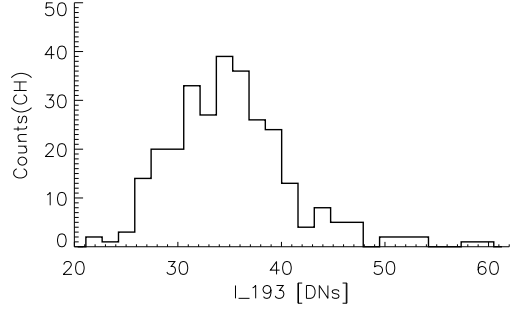}
\caption{Distribution of the mean AIA-193 intensities derived from all coronal holes.}
\label{statistics1.5}
\end{figure*}

Figure \ref{statistics1.5} shows the distribution of the mean AIA-193 intensities $\bar{I}_{193}$ of all coronal holes under study. The mean AIA-193 intensity is \SI{36 \pm 6}{DNs}. Only \SI{3}{\percent} of all coronal holes had mean intensities $\bar{I}_{193} > \SI{50}{DNs}$; the brightest coronal hole had a mean intensity of \SI{60}{DNs}. For comparison, the most common pixel intensity of the solar disk in all the AIA-\SI{193}{\angstrom} filtergrams in our dataset, i.e.\ the quiet Sun intensity, is in the range of \SI{109 \pm 16}{DNs}, and the median Sun intensity is \SI{150 \pm 17}{DNs}.

\subsection{Magnetic Properties}
\begin{figure*}[tp]
\centering
\includegraphics[width = \textwidth]{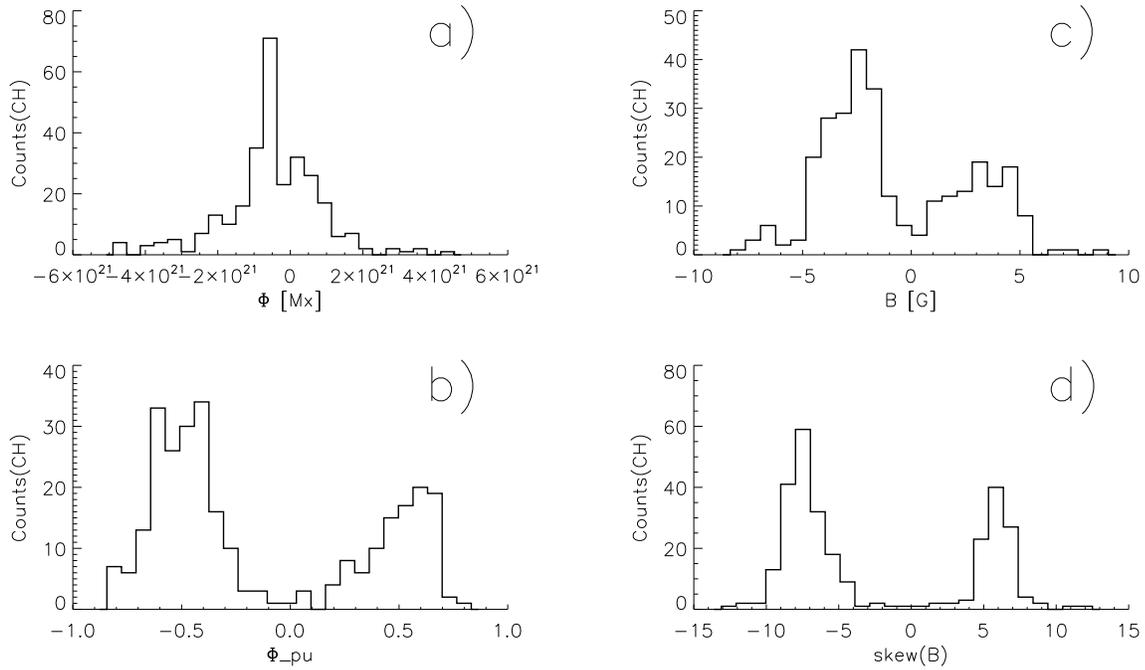}
\caption{Distribution of the unbalanced magnetic flux (a), percentaged unbalanced magnetic flux (b), and mean magnetic field densities (c), and skewness of the magnetic field distribution (d), derived from all coronal holes.}
\label{statistics2}
\end{figure*}

Figure \ref{statistics2}a and  \ref{statistics2}b show the distribution of the unbalanced magnetic flux $\Phi$ and percentaged unbalanced magnetic flux $\Phi_\text{pu}$ of coronal holes. The absolute unbalanced magnetic flux $\abs{\Phi}$ ranges from \SIrange{2.2e19}{4.7e21}{Mx} with a mean value of  $9.9 \pm 9.4 \cdot 10^{20}$~Mx. The percentaged unbalanced magnetic flux $\Phi_\text{ro}$ shows two broad peaks at $-0.46 \pm 0.15$ and $0.52 \pm 0.17$, implying that on average \SI{\approx 49}{\percent} of the magnetic flux piercing through the photosphere is not closed within the coronal hole. \SI{3}{\percent} of all coronal holes had absolute values of the percentaged unbalanced magnetic flux $\abs{\Phi_\text{pu}} > 0.75
$, and \SI{2}{\percent} of all coronal holes had $\abs{\Phi_\text{pu}} < 0.10$.

Figure \ref{statistics2}c and \ref{statistics2}d show the distribution of the mean magnetic field densities $\bar{B}$, and the skewness $\text{skew}(B)$ of the magnetic field distributions of coronal holes. The mean value of $\abs{\bar{B}}$ is \SI{2.97 \pm 1.55}{G}. The mean value of $\abs{\text{skew}(B)}$ is $6.6 \pm 1.6$, indicating that in general coronal holes have a distinct dominant polarity.

\subsection{Correlation of Characteristic Properties}

Figure \ref{statistics0}a, \ref{statistics0}b, and \ref{statistics0}c show the mean magnetic field densities $\bar{B}$ of the coronal holes versus their areas $A$, the mean AIA-193 intensities $\bar{I}_{193}$ versus the areas $A$, and the mean AIA-193 intensities $\bar{I}_{193}$ versus the mean magnetic field densities $\bar{B}$. These three parameters do not correlate. However, we note that there is some difference between small and large coronal holes, which is discussed in Section \ref{sectdiffsmalllarge}.

Figure \ref{statistics0}d shows the unbalanced magnetic flux $\Phi$ of coronal holes versus their areas $A$. The absolute values of the unbalanced magnetic flux correlate with the areas with a Pearson correlation coefficient of 0.87, the dependency can be expressed as 
\begin{equation}
\abs{\Phi} [\text{Mx}] = \num{-4.26e19} + \num{3.45e10} \cdot A \ [\text{km}^2] \label{fit_APHI}
\end{equation}
with a root mean square error (RMSE) of \num{4.79e20}{Mx}. However, \SI{66}{\percent} of the coronal holes are rather small and have an absolute value of unbalanced magnetic flux of less than \SI{1e21}{Mx}. Therefore, this relationship is not suited to predict the unbalanced magnetic flux of a coronal hole, but only to estimate the magnitude of the unbalanced magnetic flux.

Figure \ref{statistics22}a shows the mean unsigned magnetic flux $B_\text{us}$ of coronal holes versus the absolute values of their mean magnetic field densities. A distinct, linear correlation is apparent, with a Pearson's correlation coefficient of $0.976$. The relationship can be expressed as
\begin{equation}
B_\text{us} [\text{G}] = 3.17 + 0.86  \abs{\bar{B}}, \label{eq_b_babs}
\end{equation}
with a RMSE of \SI{0.30}{G}. 

Figure \ref{statistics22}b and \ref{statistics22}c show the percentage unbalanced magnetic flux of coronal holes versus the absolute values of the mean magnetic field densities and versus the mean unsigned magnetic field densities. Also here distinct relationships are apparent, with Spearman's rank correlation coefficients of $0.983$ and $0.920$. The relationships can be expressed as
\begin{align}
&\abs{\Phi_\text{pu}} = \frac{\abs{\bar{B}}\ [\text{G}]}{3.17 + 0.86 \abs{\bar{B}}\ [\text{G}]}, \label{eq_phipu_b} \\
&\abs{\Phi_\text{pu}} = \frac{-3.69 + 1.17 \  \bar{B}_\text{us}\ [\text{G}]}{\bar{B}_\text{us}\  [\text{G}]} \label{eq_phipu_babs},
\end{align}
with RMSEs of $0.025$, respectively $0.064$.
Since the percentage unbalanced magnetic flux $\Phi_\text{pu}$ can be approximated by the ratio of the mean magnetic field density to the mean unsigned magnetic field density $\bar{B}/\bar{B}_\text{us}$, Equations \ref{eq_phipu_b} and \ref{eq_phipu_babs} are directly related to Equation \ref{eq_b_babs}. Note that Equation \ref{eq_b_babs} implies that the mean magnetic field density of a coronal hole strongly depends on its mean unsigned magnetic field density, and Equation \ref{eq_phipu_b} that the amount of unbalanced magnetic flux relative to the total unsigned flux inside a coronal hole can be predicted purely by the mean magnetic field density!

\subsection{Differences in the Characteristic Properties of Small and Large Coronal Holes}
\label{sectdiffsmalllarge}
\begin{figure*}[tp]
\centering
\includegraphics[width = \textwidth]{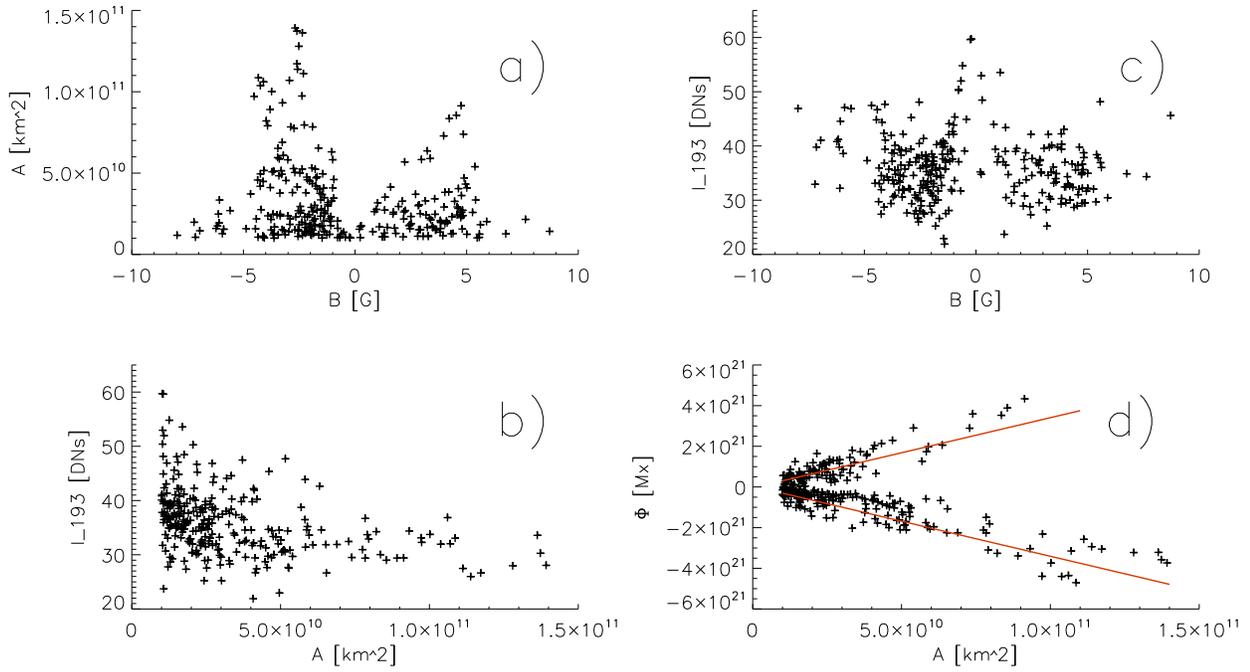}
\caption{Scatter plots of the areas versus the mean magnetic field densities (a), the mean AIA-193 intensities versus the areas (b), the mean AIA-193 intensities versus the mean magnetic field densities (c), and the unbalanced magnetic flux versus the areas (d), derived from all coronal holes. In panel d, the linear fit given by Eq. \ref{fit_APHI} is over-plotted in red.}
\label{statistics0}
\end{figure*}

Figures \ref{statistics0}a, b, and d show differences in the distribution of characteristic properties of small coronal holes, which we define as the 115 coronal holes with areas $A < \SI{2e10}{km^2}$, and the 12 large coronal holes with $A > \SI{1e11}{km^2}$. 

Small coronal holes have a wide range of mean AIA-193 intensities $\bar{I}_{193}$ ranging from \SIrange{24}{60}{DNs} (Fig.\ \ref{statistics0}b). The mean magnetic field densities $\abs{\bar{B}}$ range from \SIrange{0.2}{8.7}{G} (Fig.\ \ref{statistics0}a). The absolute values of the percentaged unbalanced magnetic flux $\abs{\Phi_\text{pu}}$ takes values of 0.06 to 0.81, i.e.\ the magnetic flux piercing through the photosphere can be balanced well as well as be almost unipolar. The absolute unbalanced magnetic flux $\abs{\Phi}$ is comparably small, ranging from \SIrange{2.2e19}{1.3e21}{Mx} (Fig.\ \ref{statistics0}d). Note that we cannot state if the small coronal holes with an almost balanced magnetic flux are coronal holes at their birth/end, or if they are artefacts.

In contrast large coronal holes always appear as dark structures with $\bar{I}_{193} < \SI{37}{DNs}$ (Fig.\ \ref{statistics0}b). They have neither very small nor very large magnetic field densities, $\abs{\bar{B}}$ ranges from \SIrange{2.3}{4.3}{G} (Fig.\ \ref{statistics0}a). The absolute values of the percentaged unbalanced magnetic flux $\abs{\Phi_\text{pu}}$ ranges from $0.40$ to $0.64$, and they always have an absolute unbalanced magnetic flux $\abs{\Phi} >  \SI{2.5e21}{Mx}$. This findings show that the characteristics of large coronal holes are much more constrained than the characteristics of small coronal holes.

\section{PIXEL DISTRIBUTION INSIDE THE LOW-LATITUDE CORONAL HOLES}
\label{pxdist}
In this section, we present the distribution of pixels \textit{inside} coronal holes. First  we derive the distribution of AIA-193 and magnetogram pixels for each coronal hole separately, and then stack the normalized distributions of all coronal holes with equal weight to obtain the median representation and the variance of the distribution of pixels. Note that this kind of representation represents the probability distributions. In particular, we present the distribution of the AIA \SI{193}{\angstrom} intensity inside coronal holes, the distribution of AIA \SI{193}{\angstrom} intensity along the major axis of coronal holes, the distribution of the magnetic field density inside coronal holes, the magnetic flux imbalance at various magnetic levels, and the percentage of the unbalanced magnetic flux that arises from various magnetic levels.

\subsection{AIA-193 Intensity}
Figure \ref{statistics11}a shows the stacked AIA-193 intensity histograms of coronal holes. The solid line gives the median AIA-193 intensity histogram, the dashed line the $1\sigma$ range, i.e.\ the range which contains \SI{68}{\percent} of all AIA-193 intensity histograms. The AIA-193 intensity is distributed broadly inside coronal holes, ranging from about \SIrange{15}{70}{DNs}; the mode is at \SI{27}{DNs}. \SI{1}{\percent} of all pixels have intensities $I_{193,i} <  \SI{16}{DNs}$,  \SI{10}{\percent} have intensities  $I_{193,i} < \SI{20}{DNs}$, \SI{10}{\percent} have intensities  $I_{193,i} > \SI{50}{DNs}$, and \SI{1}{\percent} have intensities  $I_{193,i} > \SI{65}{DNs}$. 

\subsection{AIA-193 Intensity along the Major Axis}
\begin{figure*}[tp]
\centering
\includegraphics[width = .5\textwidth]{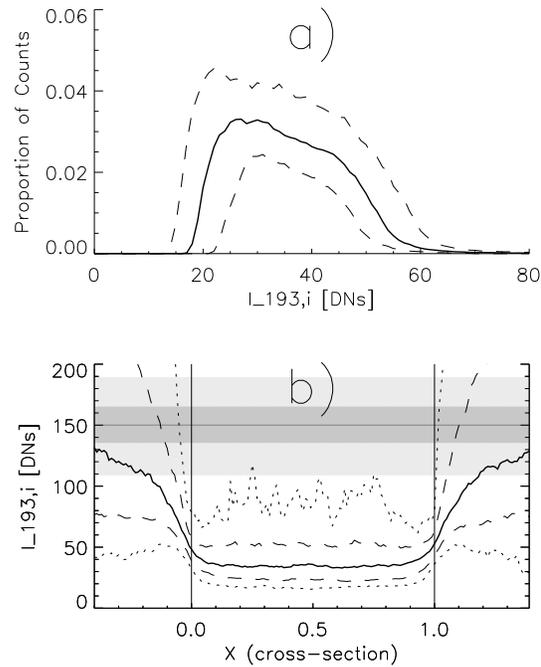}
\caption{Stacked histograms of the AIA-193 intensities of the pixel distribution of coronal holes (a), and stacked distributions of the AIA-193 intensities along the normalized cross-sections $X$ of coronal holes (b). The solid lines give the median representation, the dashed line the $1\sigma$-range, and the dotted lines the $2\sigma$ range. The straight vertical lines at $X=0$ and $X=1$ in panel b represent the borders of the coronal holes as defined by our segmentation algorithm. The grey horizontal stripes give the median value, $1\sigma$ range, and $2\sigma$ range of the median AIA-193 intensities of the entire solar disks.}
\label{statistics11}
\end{figure*}

Figure \ref{statistics11}b shows the stacked AIA-193 intensity distributions along the major axis of the coronal holes. For comparison, the horizontal stripes give the distribution of the median AIA-193 intensities of the entire solar disks (\SI{150 \pm 17}{DNs}), which is also an estimate on the median AIA-193 intensities of quiet Sun regions.
At the border of coronal holes, the median representation of the intensities steeply decreases from the quiet Sun intensity to a low intensity of about \SI{40}{DNs}. After the steep decrease, the median AIA-193 intensity stays flat, i.e.\ coronal holes do generally not reveal an intensity minimum at their center.
We further investigated the AIA-193 intensity distribution by inspecting the AIA-193 intensity distribution along the cross-section for each coronal hole separately. We found a wide variety on intensity distributions: from almost homogeneous intensity distributions without distinct intensity minimum, intensity distributions with large fluctuations which most probably arise from coronal bright points located inside of coronal holes, intensity distributions  with one and more distinct intensity minima at different locations at the cross-section, intensity distributions which rise sharply at the border of coronal holes, to intensity distributions which rise slowly and steadily at the borders. 
Thus, Figure \ref{statistics11}b represents only the statistical cross-section of the AIA-193 intensity as derived by the superposed epoch analysis, but does not hold as prototype cross-section of coronal holes.

\subsection{Magnetic Field Density}
Figure \ref{statistics4}a shows the stacked histograms of the magnetic field density of coronal holes. 
Positive values of $B_+$ represent pixels with the same polarity as the dominant polarity of the coronal hole, negative values of $B_+$ pixels with the opposite polarity. The imbalance of the magnetic flux within coronal holes is clearly visible.

\subsection{Polarity Dominance at Given Magnetic Levels}
Figure \ref{statistics4}b shows the stacked polarity dominance at given magnetic field density levels $l$, $\Phi_\text{pd}(l)$ for all coronal holes. The polarity dominance gives the fraction of the magnetic flux arising from pixels with magnetic field density $l$ which has the same polarity as the overall coronal hole. Therefore, a value of $\Phi_\text{pd}(l) = 0.5$ means that the magnetic flux arising from pixels with a magnetic field density $l$ is balanced; a value of $\Phi_\text{pd}(l)=1$ ($\Phi_\text{pd}(l)=0$) means that the magnetic flux arising from the magnetic level $l$ is unipolar and has the same (opposite) polarity as the overall coronal hole.

The median representation of the polarity dominance at given magnetic level $l$ is monotonically increasing with the magnetic level $l$; at a magnetic level of \SI{1}{G}, \SI{53}{\percent} of the magnetic flux has the same polarity as the overall coronal hole, at a magnetic level of \SI{10}{G} \SI{70}{\percent} of the magnetic flux, and at a magnetic level of \SI{30}{G} \SI{92}{\percent} of the magnetic flux. The $2\sigma$ range shows that even at low magnetic field density levels the polarity dominance is usually $> 0.5$. 
This indicates that all magnetic levels of magnetic field density predominantly have the same polarity as the overall coronal hole. The magnetic flux arising from pixels with a weak magnetic field density is only slightly unbalanced towards the dominant polarity of the coronal hole, whereas the imbalance strongly increases with increasing magnetic field density, reaching a value of $\Phi_\text{pd}(30) = 0.92$ at a magnetic field density level of \SI{30}{G}.

\subsection{Origin of the Unbalanced Magnetic Flux}

Figure \ref{statistics12}a shows the stacked cumulative histograms of the absolute values of the magnetic field density $|B_+|$ of coronal holes. In the median representation \SI{90}{\percent} of all pixels within a coronal hole have an absolute magnetic field density $\abs{B_+} < \SI{11}{G}$, \SI{9}{\percent} an absolute magnetic field density $\abs{B_+}$ of \SIrange{11}{56}{G}, and \SI{1}{\percent} an absolute magnetic field density $\abs{B_+} > \SI{56}{G}$.

Figure \ref{statistics12}b shows the stacked plots of the fraction of the unbalanced magnetic flux  $\Phi_{\text{fu}}(b)$ which arises from pixels above a given threshold magnetic field density $b$ versus $b$. In the median representation \SI{81}{\percent} of the unbalanced magnetic flux arises from pixels with an absolute magnetic field density $\abs{B_+} > \SI{10}{G}$, \SI{54}{\percent} from pixels with $\abs{B_+} > \SI{30}{G}$, and \SI{38}{\percent} from pixels with $\abs{B_+} >\SI{50}{G}$.
Therefore most of the unbalanced magnetic flux of coronal holes arises from a small percentage of its area. This aspect is studied in further detail in the following section.

\section{MAGNETIC FLUX TUBES}
\label{flxtb}

\begin{figure*}[t]
\centering
\includegraphics[width = \textwidth]{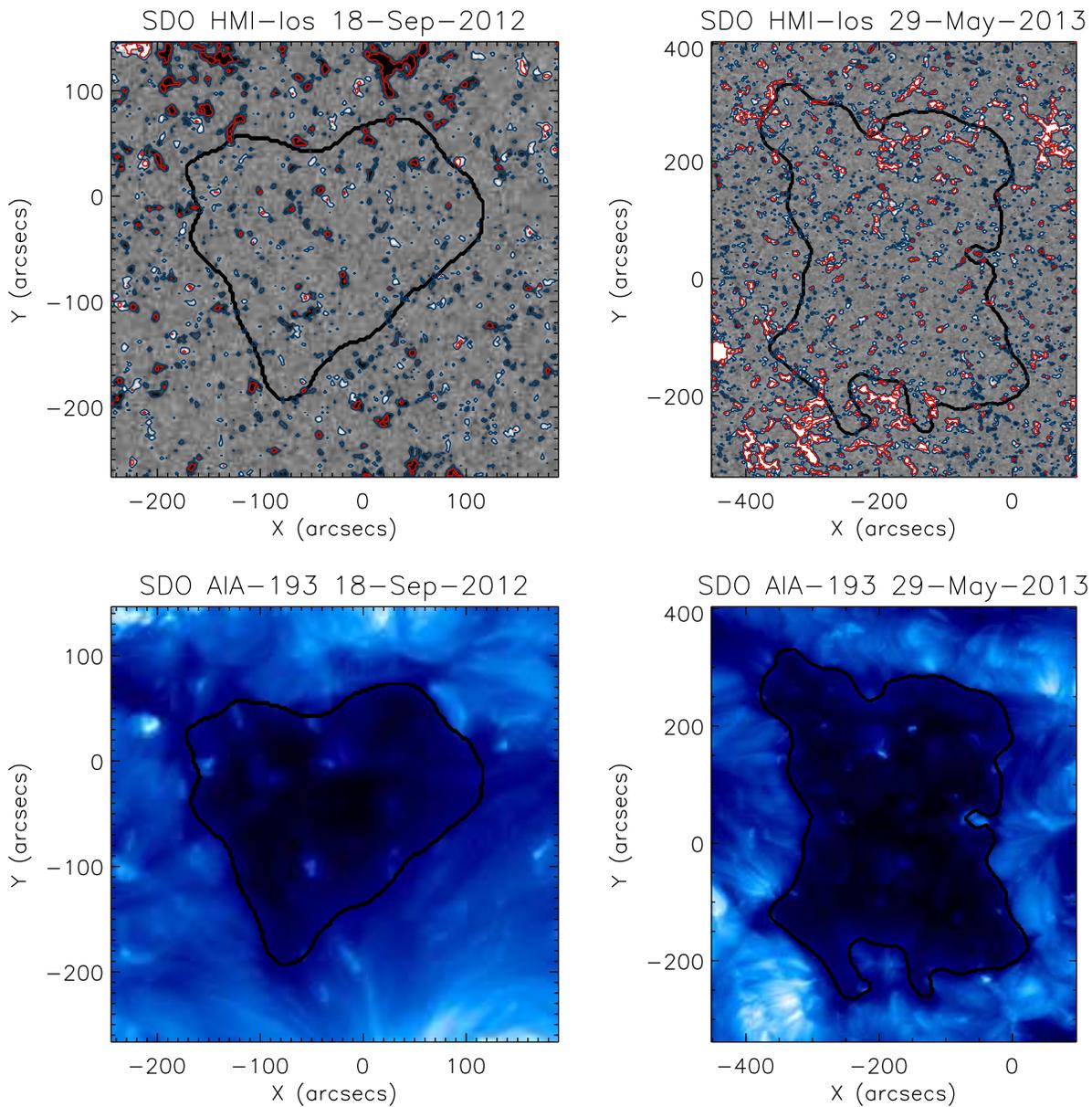}
\caption{Magnetogram (scaled to \SI{\pm 30}{G}) and \SI{193}{\angstrom} filtergram of a small sample coronal hole recorded on September 18, 2012 (left), and of a large sample coronal hole recorded on May 29, 2013 (right). The coronal hole boundaries are outlined. In the magnetogram, regions with an absolute magnetic field density of more than \SI{10}{G} are outlined in blue and of more than \SI{50}{G} in red.}
\label{ch_image_fluxtubes}
\end{figure*}

In Figure \ref{ch_image_fluxtubes}, we show two samples, a small and a large sample coronal hole: the AIA \SI{193}{\angstrom} emission together with the coronal hole boundary determined (lower panel) and the corresponding line-of-sight magnetic field map from HMI (upper panel). We find that the pixels with a high magnetic field density, i.e.\ the regions where most of the magnetic flux comes from, are clustered. We interpret these clusters as magnetic flux tubes.

In this section, we analyse the distribution of the magnetic flux and areas of all flux tubes in the coronal holes under study. 
We count the numbers of flux tubes per area per coronal hole, calculate their summed unbalanced magnetic flux and area per coronal hole, and estimate the percentage of their summed unsigned magnetic flux which is unbalanced.

Note that the flux tubes were extracted by a thresholding technique applied to the magnetograms; therefore, the quantitative parameters depend on the actual extraction threshold.	
However, the qualitative results were found to be valid for various thresholds ranging from \SIrange{10}{80}{G}. At thresholds $> \SI{20}{G}$, all flux tubes extracted in the coronal holes were unipolar, i.e.\ all pixels within the flux tube had the same polarity. Thus, the assumption of magnetic flux tubes is justified.
In the following, we present the results for flux tubes extracted at thresholds $thr$ of \SI{10}{G}, \SI{30}{G}, and \SI{50}{G}.

\subsection{Magnetic Distribution of Flux Tubes}
In total we extracted \num{26409} flux tubes at \SI{10}{G}, \num{8850} flux tubes at \SI{30}{G}, and \num{4941} flux tubes at \SI{50}{G}. 

\begin{figure*}[tp]
\centering
\includegraphics[width = \textwidth]{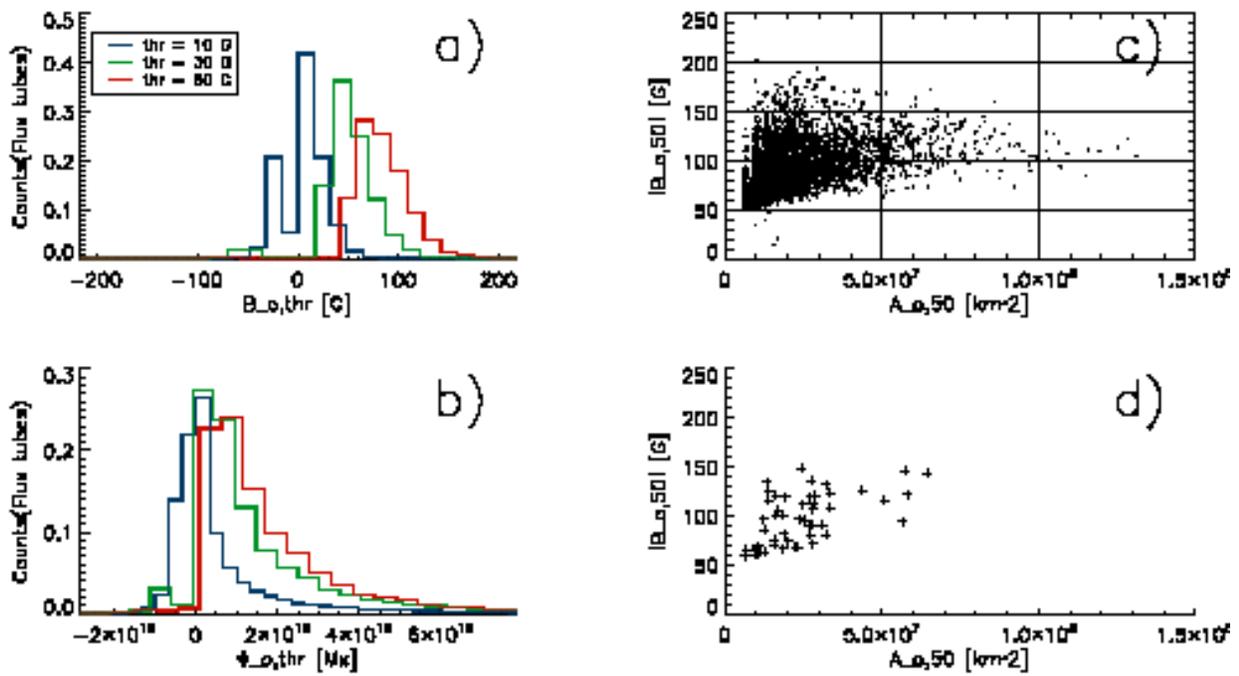}
\caption{Distribution of the mean magnetic field densities (a) and magnetic flux (b), derived from all flux tubes extracted at \SI{10}{G} (blue), \SI{30}{G} (green), and \SI{50}{G} (red). Positive values belong to the dominant polarity of their source coronal hole. Scatter plot of the mean magnetic field density of all flux tubes versus their areas derived from all flux tubes extracted at \SI{50}{G} (c), and derived for a single medium sized coronal hole (apparent on 2013/11/09; d).}
\label{statistics13}
\end{figure*}

Figure \ref{statistics13}a shows the histogram of the mean magnetic field densities $\bar{B}_{\circ,thr}$ of all flux tubes. Positive values correspond to flux tubes which have the same polarity as the dominant polarity of its coronal hole.
At thresholds of \SI{10}{G} the mean magnetic field density is \SI{18 \pm 20}{G}, at thresholds of \SI{30}{G} \SI{58 \pm 29}{G}, and at thresholds of \SI{50}{G} \SI{91 \pm 27}{G}. The positive values of the mean magnetic field densities, especially at the higher extraction thresholds, show that most flux tubes have the same dominant polarity as their coronal hole.
The lower values of mean magnetic field densities at lower extraction thresholds arises from the higher number of weak magnetic flux tubes and the larger area per flux tube.

Figure \ref{statistics13}b shows the distribution of the magnetic flux $\Phi_{\circ,thr}$ of all flux tubes. Most of the flux tubes have a small absolute magnetic flux: \SI{75}{\percent} have $\abs{\Phi_{\circ,10}} < \SI{1.0e19}{Mx}$ at a flux tube threshold of \SI{10}{G}, \SI{75}{\percent} have $\abs{\Phi_{\circ,30}} < \SI{2.2e19}{Mx}$ at a threshold of \SI{30}{G}, and \SI{75}{\percent} have $\abs{\Phi_{\circ,50}} < \SI{2.6e19}{Mx}$ at an threshold of \SI{50}{G}. However, there is also a small number of flux tubes with a large magnetic flux, with values as high as \SI{5.3e20}{Mx} at a threshold of \SI{10}{G}, \SI{2.6e20}{Mx} at a threshold of \SI{30}{G}, and \SI{1.6e20}{Mx} at a threshold of \SI{50}{G}. 

\subsection{Relationship between Area and Magnetic Field Density of Flux Tubes}

Figure \ref{statistics13}c shows the absolute mean magnetic field densities $\abs{\bar{B}_{\circ,50}}$ of all flux tubes extracted at \SI{50}{G} versus the flux tube areas $A_{\circ,50}$. \SI{55}{\percent} of all flux tubes have a small area $A_{\circ,50} < \SI{2.5e7}{km^2}$ and a small mean magnetic field density $\abs{\bar{B}_{\circ,50}} < \SI{100}{G}$. Flux tubes with a large area always have a medium to high mean magnetic field density. However, flux tubes with a very high mean magnetic field density usually have a small to medium area.

We further investigated the relationship between the absolute mean magnetic field density of flux tubes and their area extracted at \SI{50}{G} for each coronal hole individually. Figure \ref{statistics13}d shows the scatter plot of the absolute mean magnetic field densities $\abs{\bar{B}_{\circ,50}}$ of flux tubes of a typical medium sized coronal hole versus their areas $A_{\circ,50}$. At medium size and large size coronal holes a wide-spread relationship between the absolute mean magnetic field densities and the areas is often visible, with larger areas at higher mean magnetic field densities. This relationship can be explained by magnetic pressure: higher mean magnetic field densities are related to higher magnetic pressures within the flux tubes, and thus to statistically larger cross-sections.
At small coronal holes there are usually not enough flux tubes to make a meaningful statement on the correlation.

\subsection{Number of Flux Tubes per Coronal Hole}
Figure \ref{statistics19}a shows the stacked plots of the number of flux tubes per coronal hole area, $\tilde{N}_{\circ,thr} / A$, versus the extraction threshold $thr$. Coronal holes have on average $28 \pm 3$ flux tubes extracted at \SI{10}{G} per \SI{e10}{km^2}, $9 \pm 4$ flux tubes extracted at \SI{30}{G} per \SI{e10}{km^2}, and $5 \pm 3$ flux tubes extracted at \SI{50}{G} per \SI{e10}{km^2}.

\subsection{Flux Tubes as the Origin of Unbalanced Magnetic Flux}
\label{summedunbalancedmagneticflux}
In this section, we presume that magnetic flux tubes with the same polarity as the dominant polarity of the coronal hole do not close with quiet regions inside of the coronal hole, but only with flux tubes of opposite polarity inside of the coronal hole. This assumption is supported by the fact that coronal holes usually have the same dominant polarity at all magnetic scales (see Fig.\ \ref{statistics4}b), and that therefore the quiet coronal hole regions have on average also the same dominant polarity as the flux tubes with dominant polarity. In this case, the percentaged unbalanced magnetic flux of the flux tubes $\Phi_{\text{pu,thr}}$ gives a \textit{lower} estimate on the fraction of the magnetic flux arising from flux tubes which is unbalanced within the coronal hole boundaries. 

Figure \ref{statistics19}b shows the stacked plots of the percentaged unbalanced magnetic flux $\tilde{\Phi}_\text{pu,thr}$ of the flux tubes per coronal hole versus the extraction threshold. The median line shows that for \SI{50}{\percent} of all coronal holes at least \SI{85}{\percent} of the magnetic flux arising from flux tubes is unbalanced at an extraction threshold of \SI{10}{G}, at least \SI{98}{\percent} is unbalanced at an extraction threshold of \SI{30}{G}, and the complete magnetic flux is unbalanced at an extraction threshold of \SI{50}{G}. The bottom $1\sigma$ line shows that for \SI{82}{\percent} of all coronal holes at least \SI{72}{\percent} of the magnetic flux arising from flux tubes is unbalanced at an extraction threshold of \SI{10}{G}, at least \SI{93}{\percent} at \SI{30}{G}, and all the complete magnetic flux is unbalanced at \SI{50}{G}. This means that in general most of the magnetic flux tubes in coronal holes, and in particular the strong magnetic flux tubes, are open \footnote{However, the bottom $2\sigma$ line shows that there is a small percentage of coronal holes with a small percentaged unbalanced magnetic flux of flux tubes, i.e.\ where a significant fraction of the flux tubes are closed.}.

Note that Figure \ref{statistics12} can also be interpreted in the context of flux tubes.
Figure \ref{statistics12}a is proportional to the fraction of the coronal hole area that is covered by flux tubes extracted at a threshold $thr=b$, $\tilde{A}_{\circ,thr} / A$. Figure \ref{statistics12}b then shows the fraction of magnetic flux to the unbalanced magnetic flux that arises from flux tubes, $\tilde{\Phi}_{\circ,{thr}} / \Phi$. The flux tubes extracted at thresholds of \SI{10}{G} cover \SI{10}{\percent} of the area of coronal holes, and carry \SI{81}{\percent} of the unbalanced magnetic flux. If we rise the threshold to \SI{30}{G}, the flux tubes cover \SI{2.5}{\percent} of the area of coronal holes, and carry \SI{54}{\percent} of the unbalanced magnetic flux. At extraction thresholds of \SI{50}{G} the flux tubes cover \SI{1.2}{\percent} of the area of coronal holes, and carry \SI{38}{\percent} of the unbalanced magnetic flux.

This means that most of the unbalanced magnetic flux of coronal holes arise from very localized regions, i.e.\ magnetic flux tubes, covering only a small fraction of the total coronal hole area.

\section{RELATIONSHIP BETWEEN THE PROPERTIES OF LOW-LATITUDE CORONAL HOLES AND THEIR AVERAGED FLUX TUBE PARAMETERS}
\label{correlateftch}
In this section we relate the averaged parameters of magnetic flux tubes per coronal hole (cf. Sect. \ref{flxtb}) to the properties of the coronal holes (cf. Sect. \ref{stchardis}). Here we only consider flux tubes extracted at a high threshold of \SI{50}{G} (Fig. \ref{statistics14}-\ref{statistics21}: left panels) and a low threshold of \SI{10}{G} (Fig. \ref{statistics14}-\ref{statistics21}: right panels). The results are qualitatively also valid for other extraction thresholds, whereby at lower extraction thresholds the correlation coefficients get higher and the scatter in the plots lower.

\subsection{Number of Flux Tubes}
\begin{figure*}[tp]
\centering
\includegraphics[width = \textwidth]{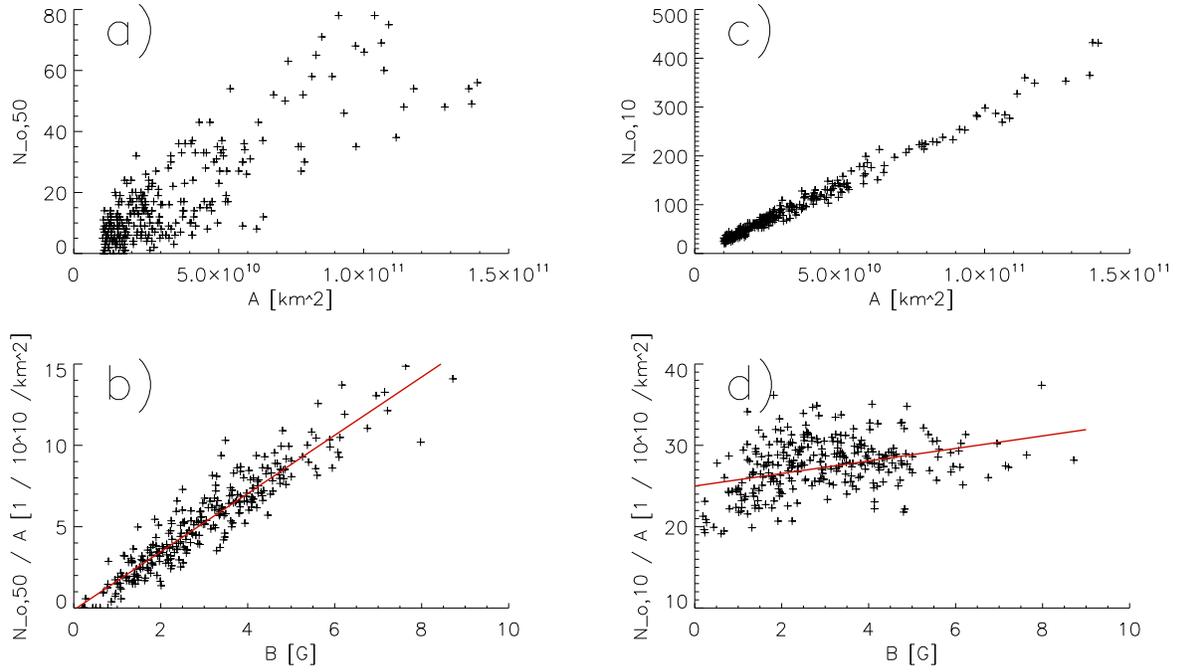}
\caption{Scatter plot of the number of flux tubes per coronal hole versus the area of the coronal hole (upper panel), and of the number of flux tubes per area versus the absolute values of the mean magnetic magnetic field density of the coronal hole (lower panel), derived for flux tubes extracted at \SI{50}{G} (left panel) and \SI{10}{G} (right panel). In panel b and d, the corresponding fits given by Eq. \ref{fitsNA50} and \ref{fitsNA10} are over-plotted in red.} 
\label{statistics14}
\end{figure*}

Figure \ref{statistics14}a shows the number of flux tubes extracted at \SI{50}{G} $\tilde{N}_{\circ,50}$ per coronal hole versus the area $A$ of coronal holes. The number of flux tubes depends linearly, but with a high scatter, on the area.
Figure \ref{statistics14}b shows the number of flux tubes per coronal hole normalized on the area of the coronal hole $\tilde{N}_{\circ,50} / A$ versus the absolute values of the mean magnetic field density $\abs{\bar{B}}$. The number of flux tubes per area depends linear on the mean magnetic field density, with about $1.5$~flux tubes per \SI{e10}{km^2} at $\abs{\bar{B}} = \SI{1}{G}$ and 9~flux tubes per \SI{e10}{km^2} at $\abs{\bar{B}}=\SI{5}{G}$. The total relationship can be expressed as 
\begin{equation}
N_{\circ,50} = \left( -0.12 + 1.79 \abs{\bar{B}} [G] \right) A [\SI{e10}{km^2}], \label{fitsNA50}
\end{equation}
with a RMSE of $2.92$ flux tubes. 

In contrast, the number of flux tubes extracted at \SI{10}{G}  $\tilde{N}_{\circ,10}$ depends mainly on the coronal hole area $A$ (Fig. \ref{statistics14}c), and only slightly on the mean magnetic field density $B$ of the coronal hole (Fig. \ref{statistics14}d), with 
\begin{equation}
N_{\circ,10} = \left( 25 + 0.77 \abs{\bar{B}} [G] \right) A [\SI{e10}{km^2}]. \label{fitsNA10}
\end{equation}
The RMSE is $10.5$ flux tubes. 

Altogether, the general number of flux tubes, i.e.\ extracted at a low extraction threshold of \SI{10}{G}, depends mainly on the area of the coronal hole, whereas the number of strong magnetic flux tubes, i.e.\ extracted at \SI{50}{G}, is also related to the mean magnetic field density of the coronal hole.

\subsection{Magnetic Field Densities}

Figure \ref{statistics20}a shows the mean magnetic field density of the quiet coronal hole regions $\bar{B}_\text{qu,50}$, i.e.\ the coronal hole without the flux tubes extracted at \SI{50}{G}, versus the averaged magnetic field density of the flux tubes extracted at \SI{50}{G} $\tilde{B}_{\circ,50}$. The mean magnetic field density of the quiet coronal hole correlates well with the averaged magnetic field density of the flux tubes, with a Spearman's rank correlation coefficient of $0.92$. However, this correlation could be artificially created by our flux tube extraction algorithm in case that the we only extracted the cores of the flux tubes, and that the outer layers of flux tubes are contained within our quiet coronal hole regions. Therefore we cross-checked this relation at low extraction thresholds of \SI{10}{G} (Fig.\ \ref{statistics20}d), and found that the correlation still holds at an Spearman's rank correlation coefficient of $0.96$. 
This findings show that strong flux tubes within the coronal hole result in a distinctly increased mean magnetic field density of the quiet coronal hole region. It also implies that the magnetic field of the flux tubes and of the quiet coronal hole area are coupled.

Figure \ref{statistics20}b shows the averaged magnetic field density derived from all flux tubes extracted at \SI{50}{G} within a coronal hole $\tilde{B}_{\circ,50}$ versus the mean magnetic field density of the coronal hole $\bar{B}$. The averaged magnetic field density of the flux tubes extracted at \SI{50}{G} correlates strongly with the mean magnetic field density of the overall coronal hole, with a Spearman's rank correlation coefficient of $0.93$. By decreasing the flux tube extraction threshold to \SI{10}{G}, the correlation coefficient even increases to $0.99$ (Fig. \ref{statistics20}e). The relationships can be expressed as 
\begin{align}
&\tilde{B}_{\circ,50} [\si{G}] = \frac{115 \ \bar{B} [\si{G}] }{ \SI{0.7}{G} + \abs{\bar{B}} [\si{G}] }, \label{fitsB50B} \\
&\tilde{B}_{\circ,10} [\si{G}] = \frac{58 \ \bar{B} [\si{G}] }{ \SI{3.9}{G} + \abs{\bar{B}} [\si{G}] }, \label{fitsB10B}
\end{align}
with a RMSE of \SI{10.7}{G} and a mean absolute percentage error (MAPE) of \SI{8.6}{\percent} for flux tubes extracted at \SI{50}{G} and a RMSE of \SI{1.6}{G} and a MAPE of \SI{6.8}{\percent} for flux tubes extracted at \SI{10}{G}. Note that these correlations result from the correlation between the average magnetic field density of flux tubes and the mean magnetic field density of the quiet coronal hole regions.

Figure \ref{statistics20}c shows the mean area per flux tube extracted at \SI{50}{G} derived from all flux tubes within a coronal hole $\tilde{A}_{\circ,50} / \tilde{N}_{\circ,50}$ versus the mean magnetic field density of the coronal hole $\abs{\bar{B}}$. 
The mean area per flux tube extracted at \SI{50}{G} correlates well with the mean magnetic field density of the coronal hole, with a Pearsons correlation coefficient of $0.64$. By decreasing the flux tube extraction threshold to \SI{10}{G}, the correlation coefficient increases to $0.93$ (Fig. \ref{statistics20}f). 
The relationships can be expressed as
\begin{align}
& \frac{ A_{\circ,50}}{ N_{\circ,50}} [\si{km^2}] = \num{1.28e7} + \num{2.13e6}  \abs{\bar{B}} [\si{G}], \label{fitsFA50B}\\
&\frac{ A_{\circ,10}}{ N_{\circ,10}} [\si{km^2}] = \num{1.58e7} + \num{6.08e6}  \abs{\bar{B}} [\si{G}] \label{fitsFA10B},
\end{align} 
with a RMSE of \SI{3.8e6}{km^2} and a MAPE of \SI{15}{\percent} for flux tubes extracted at \SI{50}{G} and a RMSE of \SI{3.6e6}{km^2} and a MAPE of \SI{8}{\percent} for flux tubes extracted at \SI{10}{G}.
These relationships can simply be explained by magnetic pressure: a high mean magnetic field density of the coronal hole is related to high magnetic field densities in the flux tubes (cf. Fig. \ref{statistics20}b and \ref{statistics20}e). A high mean magnetic field density of a flux tubes results in a high magnetic pressure within the flux tube, and therefore in a large cross-section of the flux tube under condition of comparably constant total pressure of the surrounding of the flux tube.

Altogether, the mean area per flux tube correlates well with the mean magnetic field density of the coronal hole. Furthermore, the mean magnetic field density of the coronal hole, the mean magnetic field density derived from the quiet coronal hole regions, and the averaged magnetic field density derived from all flux tubes within a coronal hole strongly correlate with each other. At lower extraction thresholds, i.e.\ by taking weaker flux tubes into account, the goodness of the relationships even enhance: the correlation coefficients increase and the MAPE decrease.

\subsection{Unbalanced Magnetic Flux}

Figure \ref{statistics21}a shows the unbalanced magnetic flux derived from all flux tubes extracted at \SI{50}{G} within a coronal hole $\tilde{\Phi}_{\circ,50}$ versus the unbalanced magnetic flux $\Phi$ of the coronal hole. The unbalanced magnetic flux of the flux tubes correlates with the unbalanced magnetic flux of the complete coronal hole, with a Spearmans rank correlation coefficient of $0.99$. By decreasing the flux tube extraction threshold to \SI{10}{G}, the Spearmans rank correlation coefficient stays at $0.99$. The relationships can be expressed as 
\begin{align}
&\tilde{\Phi}_{\circ,50} [\si{Mx}] = 0.42 \  \Phi [\si{Mx}], \label{eqflux} \\
&\tilde{\Phi}_{\circ,10} [\si{Mx}] = 0.84 \ \Phi [\si{Mx}], \label{eqflx2}
\end{align}
with a RMSE is \SI{8.9e19}{Mx} and a MAPE of \SI{79}{\percent} for flux tubes extracted at \SI{50}{G}, and a RMSE of \SI{4.7e19}{Mx} and a MAPE of \SI{7}{\percent} for flux tubes extracted at \SI{10}{G}. This means that about \SI{43}{\percent} (\SI{84}{\percent}) of the total unbalanced magnetic flux of coronal holes arises from flux tubes extracted at \SI{50}{G} (\SI{10}{G}). Note that these values slightly differ from the values derived in Section \ref{summedunbalancedmagneticflux} (\SI{38}{\percent} for flux tubes extracted at \SI{50}{G}, and \SI{81}{\percent} for flux tubes extracted at \SI{10}{G}), since the least-square fit, which was used to derive Eq. \ref{eqflux} and \ref{eqflx2}, privileges coronal holes with high magnetic fluxes. The high MAPE of Eq. \ref{eqflux} arises from the large amount of small coronal holes in our dataset, which have small magnetic fluxes and are therefore not described well by Eq. \ref{eqflux}. The MAPE of Eq. \ref{eqflx2} is significantly decreased in comparison to Eq. \ref{eqflux}, since the flux tubes extracted at \SI{10}{G} already hold most of the unbalanced magnetic flux of the overall coronal hole.

Figure \ref{statistics21}a shows the contribution of the flux tubes extracted at \SI{50}{G} on the unbalanced magnetic flux of the coronal hole, $\tilde{\Phi}_{\circ,50} / \Phi$, versus the absolute value of the average magnetic field density of the flux tubes $\abs{\tilde{B}_{\circ,50}}$. The proportion of the magnetic flux arising from the flux tubes on the unbalanced magnetic flux of the coronal hole depends on the averaged magnetic field density of the flux tubes: at a high averaged magnetic field density of all flux tubes within a coronal hole about \SI{60}{\percent} of the unbalanced magnetic flux arises from flux tubes, at a low averaged magnetic field density only about to \SI{20}{\percent}; the Spearman's rank correlation coefficient is $0.63$. By decreasing the flux tube extraction threshold to \SI{10}{G}, the amount of unbalanced magnetic flux arising from the flux tubes increases to about \SIrange{70}{87}{\percent} dependent on the averaged magnetic field density of the flux tubes; the Spearman's rank correlation coefficient is $0.62$. Taking these correlation into account, Eq. \ref{eqflux} and \ref{eqflx2} improve to
\begin{align}
&\tilde{\Phi}_{\circ,50} [\si{Mx}] = \left(-0.12 + 0.0055  \abs{\tilde{B}_{\circ,50}} [\si{G}]\right) \Phi [\si{Mx}], \label{fitsFU50B50} \\
&\tilde{\Phi}_{\circ,10} [\si{Mx}]= \left(0.72 + 0.0041  \abs{\tilde{B}_{\circ,10}} [\si{G}]\right)  \Phi [\si{Mx}], \label{fitsFU10B10}
\end{align}
with a \SI{8.4e19}{Mx} and a MAPE of \SI{30}{\percent} for flux tubes extracted at \SI{50}{G}, and a RMSE of \SI{3.2e19}{Mx} and a MAPE of \SI{4.5}{\percent} for flux tubes extracted at \SI{10}{G}.

\paragraph{}
In summary, all the relationships in this sections show that the magnetic characteristics of coronal holes are set by its magnetic flux tubes. A high mean magnetic field density of flux tubes results in a large mean area per flux tube, and thus to a high magnetic flux per flux tube and to high amount of unbalanced magnetic flux arising from flux tubes. In addition a high magnetic field density of flux tubes is related to a high mean magnetic field density of the quiet coronal hole regions, and therefore to high mean magnetic field density of the overall coronal hole.

\section{SUMMARY AND DISCUSSION}
\label{discussion}

\begin{figure*}[tp]
\centering
\includegraphics[width = .5\textwidth]{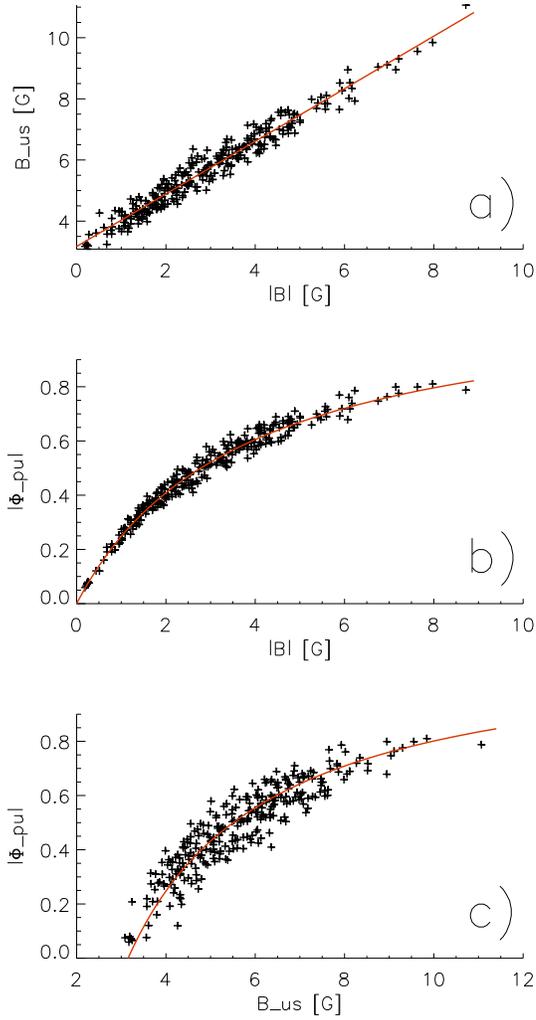}
\caption{Scatter plots of the mean unsigned magnetic field densities versus the mean magnetic field densities (a), the percentage unbalanced magnetic flux versus the mean magnetic field densities (b), and the percentage unbalanced magnetic flux versus the mean unsigned magnetic field densities (c), derived from all coronal holes. The corresponding fits given by Eq. \ref{eq_b_babs}, \ref{eq_phipu_b}, and \ref{eq_phipu_babs} are over-plotted in red.}
\label{statistics22}
\end{figure*}

\begin{figure*}[tp]
\centering
\includegraphics[width = 0.5\textwidth]{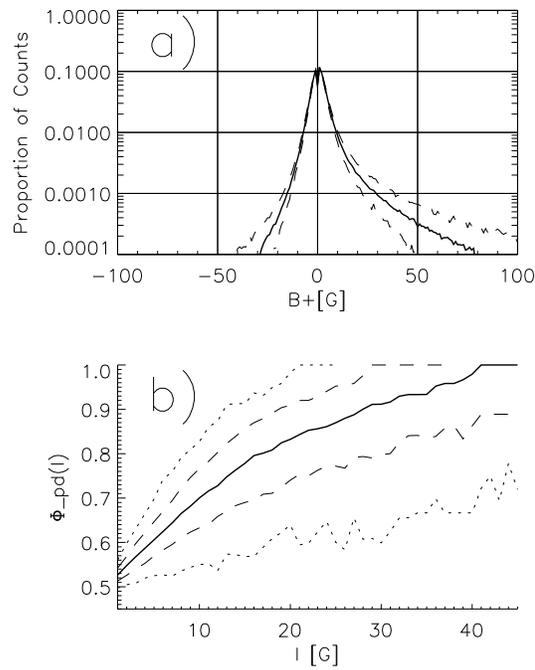}
\caption{Stacked normalized histograms of the magnetic field densities $B_+$ of coronal holes (a).
Stacked polarity dominance $\Phi_{\text{pd}}(l)$ of a given magnetic field density level $l$ versus $l$ of coronal holes (b). Positive values of $B_+$ belong to the dominant polarity of the corresponding coronal hole. The solid lines give the median values of all objects, the dashed lines the 1 $\sigma$ range, and the dotted lines the $2\sigma$ range.}  
\label{statistics4}
\end{figure*}

\begin{figure*}[tp]
\centering
\includegraphics[width = .5\textwidth]{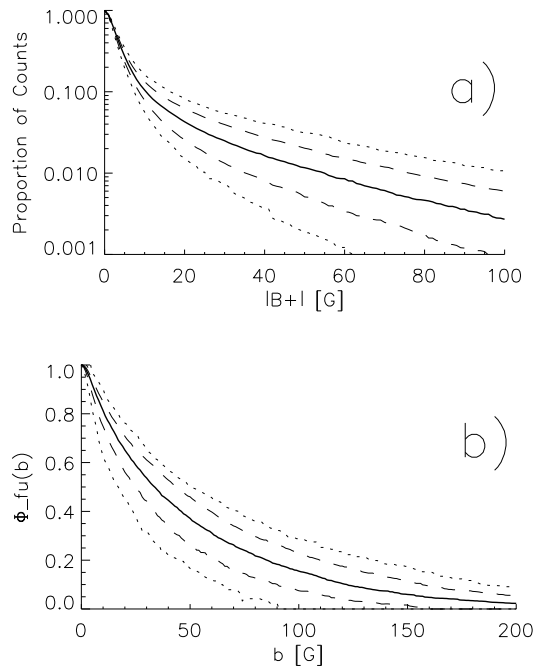}
\caption{Stacked normalized cumulative histograms of the absolute magnetic field densities $\abs{B_+}$ of coronal holes (a).
Stacked plots of the fraction of the unbalanced magnetic flux $\Phi_{\text{fu}}(b)$ arising from pixels with $\abs{B_+} > b$ on the overall unbalanced magnetic flux versus $b$ of coronal holes (b). The solid line gives the median values of all objects, the dashed lines the 1 $\sigma$ range, and the dotted lines the $2\sigma$ range.} 
\label{statistics12}
\end{figure*}

\begin{figure*}[tp]
\centering
\includegraphics[width = .5\textwidth]{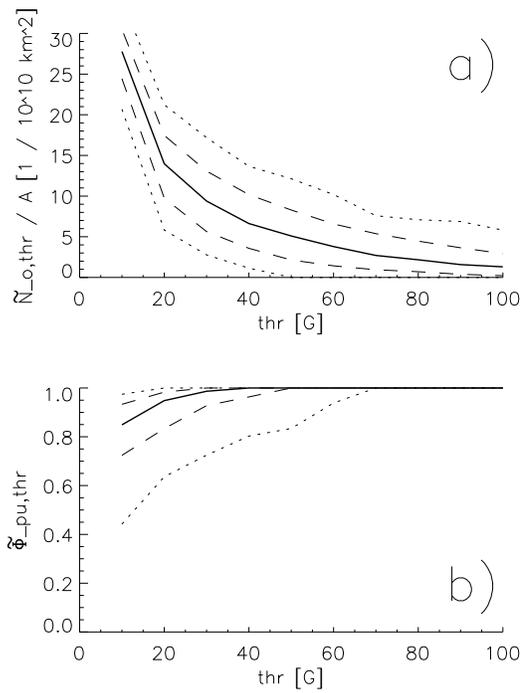}
\caption{Stacked plots of the number of flux tubes extracted at thresholds $thr$ per area versus the extraction threshold $thr$ (a). Stacked plots of the percentaged unbalanced magnetic flux of all flux tubes per coronal hole extracted at thresholds $thr$ versus the extraction threshold $thr$ (b).
The solid line gives the median values of all objects, the dashed lines the 1 $\sigma$ range, and the dotted lines the $2\sigma$ range.} 
\label{statistics19}
\end{figure*}

\begin{figure*}[tp]
\centering
\includegraphics[width = \textwidth]{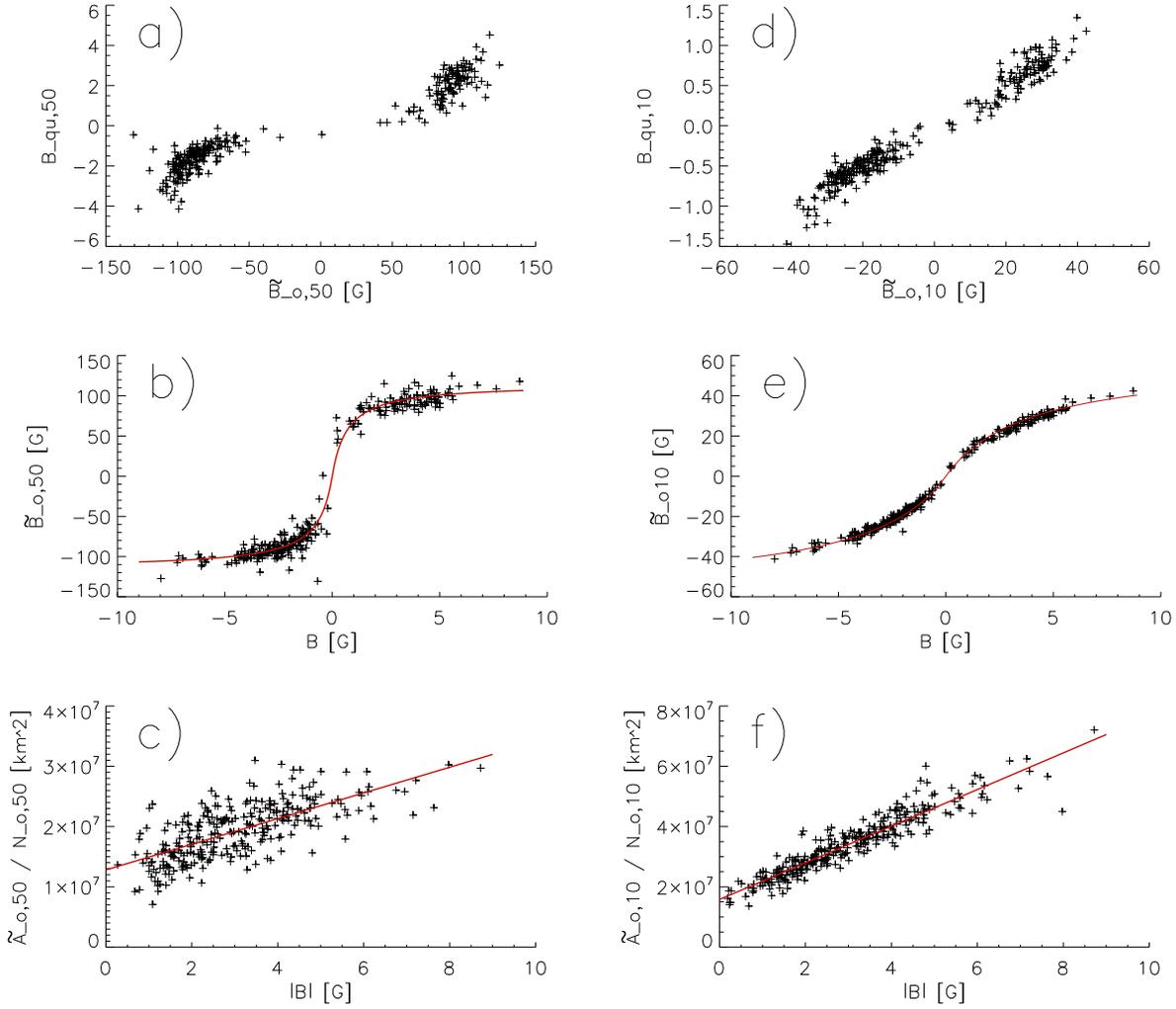}
\caption{Scatter plot of the averaged magnetic field density of all flux tubes per coronal hole versus the mean magnetic field density of the coronal hole quiet regions (upper panel), of the averaged mean magnetic field density of all flux tubes per coronal hole versus the mean magnetic field density of the coronal hole (middle panel), and of the mean area per flux tube versus the mean magnetic field density of coronal holes (lower panel), derived for flux tubes extracted at \SI{50}{G} (left panel) and \SI{10}{G} (right panel).  In panel b, c, e, and f, the corresponding fits given by Eq. \ref{fitsB50B}, \ref{fitsB10B}, \ref{fitsFA50B}, and \ref{fitsFA10B} are over-plotted in red.}
\label{statistics20}
\end{figure*}

\begin{figure*}[tp]
\centering
\includegraphics[width = \textwidth]{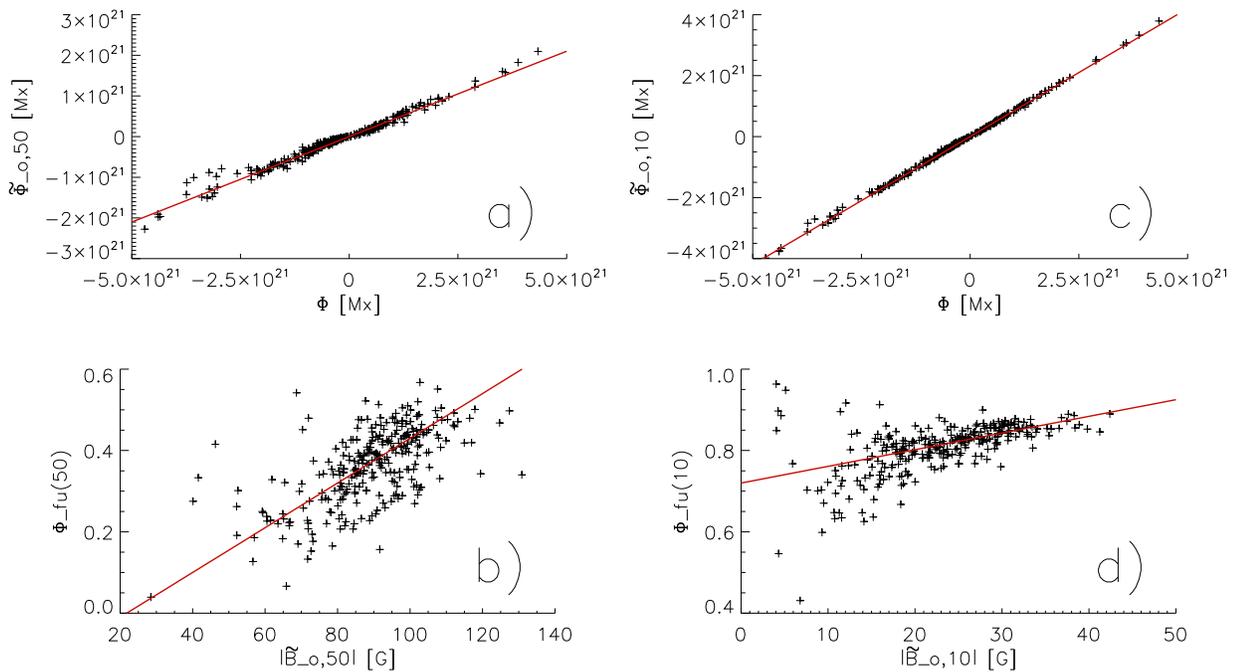}
\caption{Scatter plot of the summed unbalanced magnetic flux of all flux tubes per coronal hole versus the unbalanced magnetic flux of the coronal hole (upper panel), and of the relative amount of unbalanced magnetic flux arising from flux tubes on the unbalanced magnetic flux of coronal holes versus the averaged mean magnetic field density of flux tubes  per coronal hole (lower panel), derived for flux tubes extracted at \SI{50}{G} (left panel) and \SI{10}{G} (right panel). The corresponding fits given by Eq. \ref{eqflux}, \ref{eqflx2}, \ref{fitsFU50B50}, and \ref{fitsFU10B10} are over-plotted in red.}
\label{statistics21}
\end{figure*}

We studied the statistics of characteristic properties of 288 records of low-latitude coronal holes observed during the period 2011/01/01 to 2013/12/31, the EUV intensity and magnetic field distribution within the coronal holes, and the distribution of magnetic flux tubes. 
The main findings are:
\begin{itemize}
\item The mean magnetic field density of coronal holes ranged from \SIrange{0.2}{8.7}{G} with a mean value of \SI{3.0 \pm 1.6}{G}, the percentage unbalanced magnetic flux from \SIrange{6}{81}{\percent} with a mean value of \SI{49 \pm 16}{\percent}.
\item The mean magnetic field density, the mean unsigned magnetic field density, and the percentage unbalanced magnetic flux of coronal holes depend strongly on each other in pairs at correlation coefficients of $>0.92$ (cf. Fig. \ref{statistics22}).
\item Coronal holes have usually the same polarity on all magnetic levels: the polarity dominance at a given magnetic level $l$,  $\Phi_\text{pd}(l)$, is at all magnetic levels $l$ for almost all coronal holes always $>0.5$ (cf. Fig. \ref{statistics4}b).
\item The magnetic flux of coronal holes is concentrated in magnetic flux tubes. The flux tubes extracted at \SI{50}{G} (\SI{10}{G}) cover only \SI{1}{\percent} (\SI{10}{\percent}) of the area of coronal holes, but contain \SI{38}{\percent} (\SI{81}{\percent}) of the unbalanced magnetic flux of coronal holes (cf. Fig. \ref{statistics12}). 
\item In most coronal holes almost all magnetic flux tubes, and especially the strong magnetic flux tubes, do not close \textit{within} the coronal hole boundaries (cf. Fig.\ \ref{statistics19}b).
\item The mean magnetic field densities of coronal holes depend strongly on the averaged magnetic field density derived from all their flux tubes within the coronal hole, and the unbalanced magnetic flux of coronal holes depends strongly on the total unbalanced magnetic flux derived from all flux tubes within the coronal hole, at correlation coefficients $>0.93$ (cf. Fig. \ref{statistics20}b, e, Fig. \ref{statistics21}a, c).
\end{itemize}

The mean magnetic field densities derived in our dataset are significantly smaller than the values derived by \citet[\SIrange{3}{36}{G}]{harvey1982} and \citet[about \SI{20}{G}]{wang2009} for solar maxima, and match better with the values they derived for solar minima (\SIrange{1}{7}{G}, respectively about \SI{5}{G}). Although the selection procedure of the datasets differ -- we took all low-latitude coronal holes with areas $A > \SI{e10}{km^2}$, whereas \citet{wang2009} took all coronal holes on the solar disk, and \citet{harvey1982} took 63 selected low-latitude coronal holes -- it seems unlikely that this difference is only due to the different selection of the datasets.
 We think that the low mean magnetic field densities of our low latitude coronal holes examined near solar maximum is related to the weak polar magnetic field strength in the solar minimum before solar cycle 24 and the very low solar activity in the actual solar cycle 24 \citep[e.g.\ review by ][]{basu2013}. However, also instrumental effects of the different magnetographs used by the different authors cannot be excluded.

The characteristic parameters of coronal holes, i.e.\ the areas, the mean AIA-193 intensities, and the magnetic parameters, could further be related to the age of coronal holes. Since the area and the mean magnetic field density of coronal holes depend on the age of coronal holes \citep{bohlin1977, bohlin1978}, the unbalanced magnetic flux and the percentage unbalanced magnetic flux also depend on the age of coronal holes. 
If we assume for simplicity that the entirety of small coronal holes belong to the birth and end of coronal holes and the entirety of large coronal holes to the middle of their life time, the different range of scatter in the characteristic parameters of small to large coronal holes might simply be explained by the different phases in their life time.

We found that the average parameters of magnetic flux tubes extracted at \textit{each} extraction threshold $thr$ are strongly correlated to the magnetic parameters of the coronal hole. This implies that the average parameters of the magnetic flux tubes extracted at different thresholds $thr$ are strongly correlated, although the number of flux tubes and their amount on the unbalanced magnetic flux strongly changes with $thr$. In addition, a high average magnetic field density of flux tubes is always related to comparably high magnetic field densities of quiet coronal hole regions, i.e.\ the coronal hole region without the flux tubes. This relationship results in a distinct polarity at all magnetic levels within the coronal hole, and to a more pronounced polarity and thus a higher percentage unbalanced magnetic flux at higher mean magnetic field densities of the overall coronal hole. Thus, the magnetic characteristics of the flux tubes set the magnetic characteristics of the overall coronal hole.

The magnetic flux tubes we identified by a thresholding technique are likely to be related to the coronal magnetic funnels described in \citet{wiegelmann2005} and \citet{tu2005}. The magnetic funnels are rooted near supergranulation boundaries in the chromospheric network lanes, and are the origin of plasma outflows which form high-speed solar wind streams. The outflow velocity is statistically higher in regions with a higher magnetic field density in the underlying photosphere  \citep{xia2004}.
If we assume that our magnetic flux tubes coincide with the foot points of the magnetic funnels, then we can speculate that the averaged plasma outflow velocity might be related to the averaged magnetic field density of the magnetic flux tubes, and therefore might also be related to the mean magnetic field density of the overall coronal hole.

Finally, we note that the very strong correlation between the averaged magnetic parameters of flux tubes extracted at various magnetic levels, the quiet coronal hole regions, and the overall coronal hole parameters ($cc > 0.9$) show that there has to be an underlying mechanism which sets the distribution of flux tubes, and which relates the averaged magnetic field density of flux tubes to the mean magnetic field density of quiet coronal hole regions in an uniquely defined way. Furthermore, this mechanism has to contain the strong dependency between the mean magnetic field density, the mean unsigned magnetic field density, and the percentage unbalanced magnetic flux of coronal holes ($cc > 0.92$). This controlling mechanism is yet not known; the understanding of this mechanism might greatly improve our knowledge on coronal holes and the associated high-speed solar wind streams.

\acknowledgments
The SDO/AIA images and SDO/HMI images are available by courtesy of NASA/SDO and the AIA, EVE, and HMI science teams.
A.V. acknowledges financial support by Austrian Science Fund (FWF): P24092-N16. 
B.V. received funding from the Croatian Science Foundation under project 6212 ``Solar and Stellar Variability''.

\begin{table*}
\resizebox{.91\textwidth}{!}{
\setlength\extrarowheight{10pt}
\small
\begin{tabularx}{\linewidth}{>{$} l <{$}  >{\( \displaystyle} l <{\)} X}

\multicolumn{3}{l}{Properties of coronal holes:} \\

A  			&  = \sum_i A_i & Area   \\

\varphi 		& = \frac{1}{A} \sum_i \varphi_i A_i & Latitude of center of mass \\

I_{193} 		& = \sum_i I_{193,i} & Mean AIA-193 intensity \\

\Phi 			& = \sum_i A_i B_i  & Unbalanced magnetic flux  \\

\Phi_\text{pu} 	& = \frac{\sum_i A_i B_i}{\sum_i \abs{A_i B_i}} & percentaged unbalanced magnetic flux \\

\bar{B} 		& = \frac{\Phi}{A} & Mean magnetic field density \\

\bar{B}_\text{us} &  \frac{\sum_i \abs{A_i B_i}}{A} & Mean unsigned magnetic field density \\

\multicolumn{3}{l}{Pixel distribution within coronal holes:} \\

B_+ &=  \begin{cases} \ \ B_i & \text{for } \bar{B} \ge 0 \\ -B_i & \text{for } \bar{B} < 0 \end{cases} & Magnetic field density of a pixel, projected so that positive values belong to the dominant polarity of the coronal hole
\\[.5cm]

\Phi_\text{pd}(l) 	&= \frac{\sum_{ l - \SI{1}{G} < B_+ < l + \SI{1}{G}} B_+ A_i}{\sum_{l - \SI{1}{G} < \abs{B_+} < l + \SI{1}{G} } \abs{B_+ A_i}} & Polarity dominance of the pixels at the given magnetic field density level $l$ \\

\Phi_\text{fu}(b) 	&= \frac{1}{\Phi}\sum_{\abs{B_+} \ge b} B_+ A_i & Fraction of the unbalanced magnetic flux that arises from pixels with $B_+ \ge b$\\

\multicolumn{3}{l}{Magnetic flux tubes extracted at a threshold $thr$:} \\

A_{\circ,thr} 		& =\sum_f A_i & Area of a single flux tube \\

\Phi_{\circ,thr} 		& =\sum_f B_i A_i & Magnetic flux of a single flux tube \\

\bar{B}_{\circ,thr} 	& =\frac{\Phi_{\circ,thr}}{A_{\circ,thr}} & Mean magnetic field density of a single flux tube \\

\tilde{N}_{\circ,thr} & & Number of flux tubes of a coronal hole\\

\tilde{A}_{\circ,thr} & = \sum_{\abs{B_i} \ge thr} A_i & Total area covered by all flux tubes of a coronal hole \\

\tilde{\Phi}_{\circ,thr} & = \sum_{\abs{B_i} \ge thr} B_i A_i & Unbalanced magnetic flux derived from all flux tubes of a coronal hole \\

\tilde{\Phi}_{\text{pu,}thr} 	&= \frac{\sum_{\abs{B_i} \ge thr} B_i A_i}{\sum_{\abs{B_i} \ge thr} \abs{B_i A_i}} 		& Percentaged unbalanced magnetic flux derived from all flux tubes of a coronal hole\\

\tilde{B}_{\circ,thr}				&= \frac{1}{\tilde{A}_{\circ,thr}} \sum_{\abs{B_i} \ge thr} B_i A_i  & Average magnetic field density derived from all flux tubes of a coronal hole\\

\bar{B}_{\text{qu,}thr} 			&=  \frac{1}{A - \tilde{A}_{\circ,thr}} \sum_{\abs{B_i} < thr} B_i A_i	& Mean magnetic field density of the coronal hole region without the flux tubes\\
\end{tabularx}
}
\caption{Overview of the parameters defined in Section \ref{analysischarpars}-\ref{analysisfluxtubes}. The subscript $i$ denotes a pixel of a coronal hole, $f$ a pixel of a flux tube, $A_i$ its projection-corrected area, $B_i$ its projection-corrected mean magnetic field density, $I_{193,i}$ its mean AIA-193 intensity, and $\varphi_i$ its latitude.}
\label{DefinitionAttributes}
\end{table*}

\begin{thebibliography}{dummy}

\bibitem[Altschuler et al.(1972)]{altschuler1972} Altschuler, M.~D., Trotter, D.~E., \& Orrall, F.~Q.\ 1972, \solphys, 26, 354 

\bibitem[Basu(2013)]{basu2013} Basu, S.\ 2013, Journal of Physics Conference Series, 440, 012001 

\bibitem[Bohlin(1977)]{bohlin1977} Bohlin, J.~D.\ 1977, \solphys, 51, 377 

\bibitem[Bohlin \& Sheeley(1978)]{bohlin1978} Bohlin, J.~D., \& Sheeley, N.~R., Jr.\ 1978, \solphys, 56, 125 

\bibitem[Boucheron et al.(2016)]{boucheron2016} Boucheron, L.~E., Valluri, M., \& McAteer, R.~T.~J.\ 2016, \solphys, 291, 2353 

\bibitem[Caplan et al.(2016)]{caplan2016} Caplan, R.~M., Downs, C., \& Linker, J.~A.\ 2016, \apj, 823, 53 

\bibitem[Couvidat et al.(2016)]{couvidat2016} Couvidat, S., Schou, J., Hoeksema, J.~T., et al.\ 2016, arXiv:1606.02368 

\bibitem[Fainshtein 
\& Kaigorodov(1994)]{fainshtein1994} Fainshtein, V.~G., \& Kaigorodov, A.~P.\ 1994, \solphys, 152, 429 

\bibitem[Gnevyshev(1963)]{gnevyshev1963} Gnevyshev, M.~N.\ 1963, \sovast, 7, 311 

\bibitem[Gnevyshev(1977)]{gnevyshev1977} Gnevyshev, M.~N.\ 1977, \solphys, 51, 175 

\bibitem[Harvey et al.(1982)]{harvey1982} Harvey, K.~L., Harvey, J.~W., \& Sheeley, N.~R., Jr.\ 1982, \solphys, 79, 149 

\bibitem[Hassler et al.(1999)]{hassler1999} Hassler, D.~M., Dammasch, I.~E., Lemaire, P., et al.\ 1999, Science, 283, 810 

\bibitem[Lemen et al.(2012)]{lemen2012} Lemen, J.~R., Title, 
A.~M., Akin, D.~J., et al.\ 2012, \solphys, 275, 17 

\bibitem[Kohl et 
al.(2006)]{kohl2006} Kohl, J.~L., Noci, G., Cranmer, S.~R., \& Raymond, J.~C.\ 2006, \aapr, 13, 31 

\bibitem[Kojima et al.(2007)]{kojima2007} Kojima, M., Tokumaru, 
M., Fujiki, K., et al.\ 2007, New Solar Physics with Solar-B Mission, 369, 
549 

\bibitem[Krista 
\& Gallagher(2009)]{krista2009} Krista, L.~D., \& Gallagher, P.~T.\ 2009, \solphys, 256, 87 

\bibitem[Levine et al.(1977a)]{levine1977a} Levine, R.~H., Altschuler, M.~D., \& Harvey, J.~W.\ 1977, \jgr, 82, 1061 

\bibitem[Levine et al.(1977b)]{levine1977b} Levine, R.~H., Altschuler, M.~D., Harvey, J.~W., \& Jackson, B.~V.\ 1977, \apj, 215, 636 

\bibitem[Lowder et al.(2014)]{lowder2014} Lowder, C., Qiu, J., Leamon, R., \& Liu, Y.\ 2014, \apj, 783, 142 

\bibitem[Nolte et al.(1976)]{nolte1976} Nolte, J.~T., Krieger, 
A.~S., Timothy, A.~F., et al.\ 1976, \solphys, 46, 303 

\bibitem[P{\"o}tzi et al.(2015)]{potzi2015} P{\"o}tzi, W., 
Veronig, A.~M., Riegler, G., et al.\ 2015, \solphys, 290, 951 

\bibitem[Reiss et al.(2016)]{reiss2016} Reiss, M. A., Temmer, M., Veronig, A. M., Nikolic, L., Vennerstrom, S., Sch\"ongassner, F., Hofmeister, S. J.\ 2016, Space Weather, doi: 10.1002/2016SW001390


\bibitem[Rotter et al.(2012)]{rotter2012} Rotter, T., Veronig, 
A.~M., Temmer, M., \& Vr{\v s}nak, B.\ 2012, \solphys, 281, 793 

\bibitem[Scherrer et al.(2012)]{scherrer2012} Scherrer, P.~H., Schou, J., Bush, R.~I., et al.\ 2012, \solphys, 275, 207 

\bibitem[Schou et al.(2012)]{schou2012} Schou, J., Scherrer, P.~H., Bush, R.~I., et al.\ 2012, \solphys, 275, 229 

\bibitem[Tu et al.(2005)]{tu2005} Tu, C.-Y., Zhou, C., Marsch, E., et al.\ 2005, Science, 308, 519 

\bibitem[Verbeeck et al.(2014)]{verbeek2014} Verbeeck, C., Delouille, V., Mampaey, B., \& De Visscher, R.\ 2014, \aap, 561, A29 

\bibitem[Wang 
\& Sheeley(1990)]{wang1990} Wang, Y.-M., \& Sheeley, N.~R., Jr.\ 1990, \apj, 355, 726

\bibitem[Wang(2009)]{wang2009} Wang, Y.-M.\ 2009, \ssr, 144, 383 

\bibitem[Wiegelmann \& Solanki(2004)]{wiegelmann2004} Wiegelmann, T., \& Solanki, S.~K.\ 2004, \solphys, 225, 227 

\bibitem[Wiegelmann et al.(2005)]{wiegelmann2005} Wiegelmann, T., Xia, L.~D., \& Marsch, E.\ 2005, \aap, 432, L1 

\bibitem[Xia et al.(2004)]{xia2004} Xia, L.~D., Marsch, E., \& Wilhelm, K.\ 2004, \aap, 424, 1025 

\end{thebibliography}
\end{document}